\documentclass[prd,10pt,tightenlines,aps,letterpaper,amsmath,amssymb,preprintnumbers,showpacs,floatfix,longbibliography,nofootinbib,superscriptaddress,twocolumn]{revtex4-2}

\usepackage{amsmath}
\usepackage{amsfonts}
\usepackage{amssymb}
\usepackage{physics}
\usepackage{slashed}
\usepackage{simplewick}
\usepackage{braket}
\usepackage{hyperref}

\usepackage[utf8]{inputenc}
\usepackage[normalem]{ulem}

\usepackage{siunitx}

\usepackage{graphicx}
\usepackage{float}

\usepackage{listings}

\usepackage{tikz}
\usetikzlibrary{calc}
\usetikzlibrary{patterns}
\usetikzlibrary{patterns}
\usetikzlibrary{positioning,decorations.pathmorphing,decorations.markings}

\usetikzlibrary{decorations.pathreplacing, decorations.shapes}

\usepackage{feynmp-auto}
\usepackage{multirow}

\usepackage{xcolor}

\begin{document}
\begin{fmffile}{feynDiags}
\end{fmffile}

\begin{abstract}
The pion light-cone distribution amplitude (LCDA) is an essential non-perturbative input for a range of high-energy exclusive processes in quantum chromodynamics. Building on our previous work, the continuum limit of the fourth Mellin moment of the pion LCDA is determined in quenched QCD using quark masses which correspond to a pion mass of $m_\pi=550$ MeV. This calculation finds $\braket{\xi^2} = 0.202(8)(9)$ and $\braket{\xi^4} = 0.039(28)(11)$ where the first error indicates the combined statistical and systematic uncertainty from the analysis and the second indicates the uncertainty from working with Wilson coefficients computed to next-to-leading order. These results are presented in the $\overline{MS}$ scheme at a renormalization scale of $\mu = 2~\si{GeV}$. 
\end{abstract}

\title{Parton physics from a heavy-quark operator product expansion: Lattice QCD calculation of the fourth moment of the pion distribution amplitude}

\author{William Detmold}
\email{wdetmold@mit.edu}
\affiliation{Center for Theoretical Physics -- a Leinweber Institute, Massachusetts Institute
  of Technology,
Cambridge, MA 02139, USA}

\author{Anthony V. Grebe}
\email{agrebe@mit.edu}
\affiliation{Center for Theoretical Physics -- a Leinweber Institute, Massachusetts Institute of Technology, 
Cambridge, MA 02139, USA}
\affiliation{Fermi National Accelerator Laboratory, 
Batavia, IL 60502, USA}

\author{Issaku Kanamori}
\email{kanamori-i@riken.jp}
\affiliation{RIKEN Center for Computational Science, Kobe 650-0047, Japan}

\author{C.-J. David Lin}
\email{dlin@nycu.edu.tw}
\affiliation{Institute of Physics, National Yang Ming Chiao Tung University, Hsinchu 30010, Taiwan}
\affiliation{Centre for High Energy Physics, Chung-Yuan Christian University, Taoyuan, 32032, Taiwan}

\author{Robert J. Perry}
\email{perryrobertjames@gmail.com}
\affiliation{Center for Theoretical Physics -- a Leinweber Institute, Massachusetts Institute
  of Technology,
Cambridge, MA 02139, USA}

\author{Yong Zhao}
\email{yong.zhao@anl.gov}
\affiliation{Physics Division, Argonne National Laboratory, Lemont, IL 60439, USA}

\collaboration{HOPE Collaboration}

\maketitle

\preprint{MIT-CTP/5913, FERMILAB-PUB-25-0615-T}

\section{Introduction}
Understanding the internal structure of hadrons from the fundamental theory of the strong force, quantum chromodynamics (QCD), is a central goal of nuclear physics. Due to asymptotic freedom, certain short-distance properties of hadrons, including observables relevant for the study of their internal structure may be predicted using perturbation theory. In order to make contact with experiments, it is essential that these perturbative calculations are combined with non-perturbative hadronic matrix elements that describe long-distance physics. The separation of an observable into perturbatively calculable short-distance kernels and long-range non-perturbative matrix elements is known as a QCD factorization theorem. These theorems can be obtained for a range of physical processes including deep inelastic scattering (DIS) and the Drell-Yan process. Factorization theorems also exist for a range of high-energy exclusive processes in QCD. One of the most well-known of these is the factorization for the pion electromagnetic form factor, $F_\pi(Q^2)$ which states that at asymptotically-high $Q^2$, where $\Lambda_\text{QCD}^2/Q^2\to0$~\cite{Radyushkin:1977gp,Farrar:1979aw,Lepage:1980fj},
\begin{equation}
Q^2 F_\pi(Q^2)=16\pi f_\pi^2\alpha_S(Q^2)\,, 
\end{equation}
with $f_\pi=0.132~\si{GeV}$ being the pion decay constant and $\alpha_S(Q^2)$ being the QCD running coupling evaluated at the scale $Q^2$. The asymptotic $Q^2\to\infty$ limit clearly cannot be reached in experiment.
However, it is natural to expect that at sub-asymptotic energies, the above relation should hold modulo perturbative corrections and higher-twist contributions. 

At experimentally accessible kinematics, the more appropriate expression is
\begin{equation}
\label{eq:sub-asymptotic}
Q^2F_\pi(Q^2)=16\pi f_\pi^2\alpha_S(Q^2) w^2(Q^2)\,,
\end{equation}
which differs from the asymptotic form by the factor~\cite{Lepage:1980fj,Chang:2013nia}
\begin{equation}\label{eq:w_factor}
w(\mu) = \frac{1}{3}\int_{-1}^1 d\xi\, \frac{\phi(\xi,\mu)}{1+\xi},
\end{equation}
where $\phi(x,\mu^2)$ is the light-cone distrbution amplitude (LCDA) and plays a somewhat analogous role to the parton distribution function (PDF) for exclusive processes. In light-cone gauge, the LCDA has the interpretation of a probability amplitude for a meson (in this case, the pion) to convert into a collinear quark and anti-quark pair with momentum fractions $(1+\xi)/2$ and $(1-\xi)/2$, respectively. This transition amplitude therefore carries the non-perturbative information about the electromagnetic form factor, $F_\pi(Q^2)$. Assuming the above integral converges, it is possible to expand the denominator of the integrand and integrate term by term. Thus one finds the equivalent expression
\begin{equation}\label{eq:w_factor_series}
w(\mu) = \frac{2}{3}\sum_{n=0}^\infty (-1)^n \braket{\xi^n}(\mu),
\end{equation}
where
\begin{equation}\label{eq:mellin_moments}
\braket{\xi^n}(\mu) = \frac{1}{2} \int_{-1}^1 d\xi \xi^n \phi(\xi,\mu),
\end{equation}
defines the $n$th Mellin moment of the LCDA.

A general property of the non-perturbative long-range components of factorization theorems is \textit{process independence}. This feature increases the predictability of these theorems, since it implies the same non-perturbative inputs enter into a range of different physical observables. In the case of the pion LCDA, it can be shown that this quantity is essential for a high-energy description of decays of heavy hadrons~\cite{Dugan:1990de,Beneke:1999br,Keum:2000wi,Bauer:2005kd}, which are sensitive to CKM matrix elements. In addition, processes like deeply-virtual meson production (DVMP) admit a factorization theorem in which the pion LCDA is a required input~\cite{Collins:1996fb}.

Due to its non-perturbative character, it is natural to attempt to learn about the LCDA using lattice QCD (LQCD). Formally, the LCDA is defined as
\begin{equation}
\label{eq:pion_DA_def}
\begin{split}
\bra{ 0 } \overline{\psi}_d(z) \gamma_\mu \gamma_5 & \mathcal{W}[z, -z] \psi_u(-z) 
\ket{\pi^+ (\vec{p}) }
\\
&= i  p_\mu f_\pi \int_{-1}^1 d \xi \,
 e^{-i \xi p\cdot z }\phi_\pi(\xi, \mu ) ,
\end{split}
\end{equation}
where ${\mathcal{W}}[-z,z]$ is a light-like Wilson line ($z^{2} = 0$).  In the above equation, ${\vec{p}}$, and $p_{\mu}$ are 
the three-momentum and the four-momentum of the pion, respectively.
While matrix elements of the light-like operator defined above can in principle be directly computed using Minkowski-space formulations of lattice field theory~\cite{Drell:1978hr,Lamm:2019uyc, Echevarria:2020wct, Kang:2025xpz, Banuls:2025wiq, Chen:2025zeh}, so far it has only been feasible for lower dimensional theories such as the Schwinger model. This light-like operator poses challenges for a direct evaluation of this quantity in Euclidean-space formulations of four-dimensional LQCD. The operator product expansion (OPE) can be used to relate this non-local operator to a set of local operators, which can be directly computed using LQCD~\cite{Kronfeld:1984zv,Martinelli:1987si,Arthur:2010xf,RQCD:2019osh}. These local operators can be directly related to the Mellin moments defined in Eq.~\eqref{eq:mellin_moments}.

More recently, new ideas have led to alternate methods for calculating the LCDA and related hadronic structure quantities~\cite{Liu:1993cv,Aglietti:1998ur,Liu:1999ak,Detmold:2005gg,Braun:2007wv,Davoudi:2012ya,Ji:2013dva,Ji:2014gla,Ji:2020ect,Chambers:2017dov,Radyushkin:2017cyf,Orginos:2017kos,Ma:2017pxb,Hansen:2017mnd,Hansen:2019idp,Shindler:2023xpd,Gao:2023lny}. These other methods may be separated into inverse-problem techniques~\cite{Liu:1993cv,Liang:2019frk,Hansen:2017mnd,Hansen:2019idp}, 
OPE or short-distance-factorization approaches~\cite{Aglietti:1998ur,Liu:1999ak,Detmold:2005gg,Braun:2007wv,Davoudi:2012ya,Chambers:2017dov,Radyushkin:2017cyf,Orginos:2017kos,Ma:2017pxb,Shindler:2023xpd}, and large-momentum effective theory (LaMET)~\cite{Ji:2013dva,Ji:2014gla,Ji:2020ect,Gao:2023lny}, which permits direct calculation of the Bjorken-$x$ dependence in a limited range. 
In practice, these approaches provide complementary information on the parton distributions and can be combined to maximize the predictive power~\cite{Ji:2025hcy}.

In this work, an approach first proposed in Ref.~\cite{Detmold:2005gg} and since termed the heavy-quark OPE (HOPE) method is employed to study the low moments of the pion LCDA. In this approach, a two-current hadronic matrix element is computed using LQCD. The currents are chosen to be flavor-changing heavy-light currents. The theoretical approach was further developed in Ref.~\cite{Detmold:2021uru} and the first numerical determination of the second Mellin moment of the pion LCDA was presented in Ref.~\cite{Detmold:2021qln}. While this calculation demonstrated that the HOPE method was a viable method for studying the lowest non-trivial Mellin moment, it did not demonstrate the central benefit of the HOPE method, namely that the approach can be applied to study the higher Mellin moments. Therefore, the goal of this work is to study the prospects for a controlled calculation of the fourth Mellin moment, including the choice of kinematics and the relative sensitivity of this method to the higher moments. 

In this work, a numerical calculation of the the second Mellin moment and, for the first time, the fourth Mellin moment of the pion LCDA in the continuum limit is presented. 
After including estimated uncertainties from the effects of quenching, unphysical pion mass and finite lattice spacings present in other calculations, the results presented here agree with other results presented in the literature. In addition, the second moment agrees with the previous determination using the HOPE method~\cite{Detmold:2021qln}.  These results suggest that the HOPE method is a viable technique to learn about the higher moments of hadron structure observables like the LCDA. In particular, this numerical study demonstrates for the first time direct evidence that the HOPE method is able to sidestep the problem of power divergences present in the local operator approach, which has limited that technique to the second Mellin moment~\cite{Martinelli:1987si}. The structure of this paper is as follows: in Section~\ref{sec:hope_method} the HOPE method is introduced, in Section~\ref{sec:numerical_details} the numerical details of the study are explained and results for the second and fourth Mellin moments are presented, in Section~\ref{sec:discussion} the values determined here are compared to other results in the literature, and in Section~\ref{sec:conclusions} the conclusions of this work are summarized.

\section{The HOPE Method}\label{sec:hope_method}
The HOPE method allows one to extract information about non-perturbative quantities which feature in QCD factorization theorems from the Euclidean time dependence of hadronic matrix elements. In the case of the pion LCDA, information about the Mellin moments can be obtained from a study of the hadronic matrix element
\begin{widetext}
\begin{equation}
  R^{\mu\nu}(t_-; \vec{p}, \vec{q})
= \int d^3 \vec{z} \,e^{i\vec{q}\cdot \vec{z}} \braket{ 0 |T\{ J_A^\mu(t_-/2,\vec{z}/2) J_A^\nu(-t_-/2,-\vec{z}/2)\} | \pi(\vec{p}) },
\label{eq:ratio}
\end{equation}
\end{widetext}
where $J_{A}^{\mu}$ is an axial-vector current involving a light quark $\psi$ and a fictitious valence heavy quark $\Psi$ field given by
\begin{equation}
 J^{\mu}_{A} = \overline{\Psi} \gamma^{\mu}\gamma^{5}\psi + \overline{\psi} \gamma^{\mu}\gamma^{5} \Psi  .
\label{eq:axial_current}
\end{equation}
Due to symmetry arguments, it can be shown that this hadronic matrix element is antisymmetric under interchange of $\mu$ and $\nu$~\cite{Detmold:2021qln}. The time-momentum representation of the hadronic tensor is related to the momentum space representation via
\begin{equation}
  R^{\mu\nu}(t_-; \vec{p}, \vec{q}) = \int \frac{dq_4}{(2\pi)} \, e^{-iq_4 t_-}  V^{\mu\nu}(q,p) .
  \label{fourier-transform-temporal}
\end{equation}

The HOPE expression for the momentum space hadronic matrix element $V^{\mu\nu}(p,q)$ in the $\overline{\text{MS}}$ scheme is~\cite{Detmold:2021uru}
\begin{widetext}
\begin{equation}
\label{eq:Gegen_OPE_had_amp}
 V^{\mu\nu} (q,p) = - \frac{2 i  f_{\pi}\epsilon^{\mu\nu\rho\sigma}
 q_{\rho} p_{\sigma}}{\tilde{Q}^{2}}  
 \sum_{n=0,{\mathrm{even}}}^{\infty} {\mathcal{F}}_{n} (\tilde{Q}^{2},
 \mu, \tilde{\omega}, m_\Psi) \phi_{n} (\mu)  + 
  \text{higher-twist terms}\, ,
\end{equation}
\end{widetext}
where $m_{\Psi}$ is the mass of the fictitious valence heavy quark and ${\mathcal{F}}_{n}$  are coefficients that can be computed in QCD perturbation theory and can be expressed as functions of the kinematic variables
\begin{equation}
\label{eq:kin_var_def}
 \tilde{Q}^{2} = Q^{2} + m_{\Psi}^{2} \, , ~~~~~
 \tilde{\omega} = \frac{2 p\cdot q}{\tilde{Q}^{2}}
\end{equation}
with $Q^{2} = -q^{2}$. The explicit form of the $\mathcal{F}_n$ to one-loop order can be found in  Ref.~\cite{Detmold:2021uru}. The coefficients $\phi_{n} (\mu)$ are the Gegenbauer moments defined by the relation
\begin{equation}
\phi_n(\mu)=\frac{(2n+3)}{3(n+1)(n+2)}\int_{-1}^{1} d\xi\, \mathcal{C}_n^{3/2}(\xi)\phi(\xi,\mu),
\end{equation}
where $\mathcal{C}_n^{3/2}(\xi)$ are the Gegenbauer polynomials.
The Gegenbauer moments do not mix under the renormalization group scale evolution at one-loop order due to conformal symmetry~\cite{Efremov:1978rn}. They can be related to the Mellin moments defined above via
\begin{align}
\phi_0(\mu) &= \braket{\xi^0}(\mu) = 1,
\\
\phi_2(\mu) &= \frac{7}{12}\bigg[5\braket{\xi^2}(\mu)-\braket{\xi^0}(\mu)\bigg], 
\\
\phi_4(\mu) &= \frac{11}{24}\bigg[21\braket{\xi^4}(\mu)-14\braket{\xi^2}(\mu)+\braket{\xi^0}(\mu)\bigg].
\end{align}
The main theoretical advantage of the HOPE framework is the presence of the heavy-quark field which serves as an additional hard scale to suppress higher-twist corrections. 

\subsection{Lattice Correlation Functions}
The time-momentum representation $R^{\mu\nu}(t_-,\vec{p},\vec{q})$ can be determined from a suitable ratio of correlation functions. In particular, the three-point correlation function
\begin{figure}
\centering
\begin{tikzpicture}
\draw[line width=1.5] (0,0) .. controls (0.8,1.) and (2.2,1.4) .. (3,1.5); 
\draw[line width=4] (3,1.5) .. controls (3.8,1.4) and (5.2,1.) .. (6,0); 
\draw[line width=1.5] (0,0) .. controls (2,-2) and (4,-2) .. (6,0); 

\draw[line width=1,decorate,decoration={coil,aspect=0}] (3,1.5) -- (3,2.75);
\draw[line width=1,decorate,decoration={coil,aspect=0}] (6,0) -- (6,1.25);

\filldraw[black] (0,0) circle (4mm);
\filldraw[white] (0,0) circle (3.5mm);

\filldraw[black] (3,1.5) circle (1.5mm);
\filldraw[black] (6,0) circle (1.5mm);

\node at (0,0) {$\mathcal{O}_\pi^\dagger$};

\node at (2,1.75) {$J_A^\mu(t_e,\vec{x}_e)$};
\node at (7.2,0) {$J_A^\nu(t_m,\vec{x}_m)$};

\node at (6,1.7) {$\vec{p}_m$};
\node at (3,3.2) {$\vec{p}_e$};

\draw[->] (5.5,1.7) -- (5.5,1.0);
\draw[->] (3.5,3.2) -- (3.5,2.5);

\end{tikzpicture}
\caption{Topology of required Wick contraction for the calculation of Eq.~\ref{eq:3pt_corr}. Thin solid lines denote light quark propagators, while the thick line denotes the heavy-quark propagator. Arrows denote the convention for positive momentum.}
\label{fig:contraction}
\end{figure}
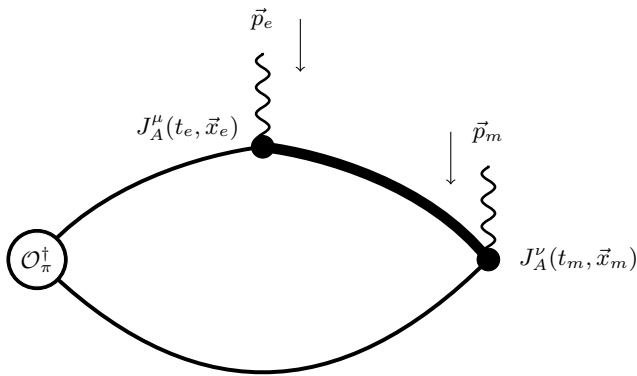
\begin{widetext}
\begin{align}
\mathcal{C}^{\mu\nu}_3 (t_e, t_m; \vec{p}_e, \vec{p}_m)
&= \sum_{\vec{x}_e\in \Lambda}\sum_{\vec{x}_m\in \Lambda} e^{i\vec{p}_e\cdot \vec{x}_e}e^{i\vec{p}_m\cdot \vec{x}_m}\braket {0 | T\{ J_A^\mu(t_e, \vec{x}_e) J_A^\nu(t_m, \vec{x}_m) \mathcal{O}^\dagger_\pi(\vec{0},0) \} | 0 } ,
\label{eq:3pt_corr}
\end{align}
\end{widetext}
can be computed using the sequential source method~\cite{Kilcup:1985fq}. The sums in the above expression are over the lattice site locations defined by the set
\begin{equation}
\Lambda = \{(\vec{n},n_4)~|~n_i = 0,1,\dots N_i-1\}
\end{equation}
where $N_i = L/a$ for $i=1,2,3$ and $N_4 = T/a$. Here and below, quantities directly computable from stochastic estimates of path integrals are represented using calligraphic lettering. The operator $\mathcal{O}_\pi(\vec{0},0)$ is an operator chosen to posess the quantum numbers of the pion. It is convenient to work with the set of coordinates 
\begin{align}
t_+ &= t_e+t_m, & t_- &= t_e - t_m, \\
\vec{p}&=\vec{p}_e+\vec{p}_m, & \vec{q}&=(\vec{p}_e-\vec{p}_m)/2.    
\end{align}
The topology of the required Wick contraction is shown in Fig.~\ref{fig:contraction}. By studying the large-$t_+$ dependence of the above correlator, it is possible to show that
\begin{equation}
\begin{split}
\lim_{t_+\to\infty}\mathcal{C}_3^{\mu\nu}&(t_e, t_m; \vec{p}_e, \vec{p}_m) 
\\
&=R^{\mu\nu}(t_-; \vec{p}, \vec{q}) \frac{Z_\pi(\vec{p})}{2E_\pi(\vec{p})}e^{-E_\pi(\vec{p})t_+/2} 
\end{split}
\end{equation}
with $\vec{p} = \vec{p}_e + \vec{p}_m$. The factor $Z_\pi(\vec{p})$ is given by
\begin{equation}
Z_\pi(\vec{p})=\braket{\pi(\vec{p})|\mathcal{O}_\pi(0)^\dagger|0}\,,
\end{equation}
while $E_\pi(\vec{p})$ is the energy-eigenvalue corresponding to the energy-eigenstate $\ket{\pi(\vec{p})}$. Both of these numbers can be extracted from a study of the two-point correlation function
\begin{equation}
\mathcal{C}_2 (t; \vec{p}) = \sum_{\vec{x}\in |\Lambda|} e^{i\vec{p}\cdot \vec{x}}\braket{0|\mathcal{O}_\pi(\vec{x},t)\mathcal{O}_\pi^\dagger(\vec{0},0)|0},
\end{equation}
which admits a spectral decomposition as
\begin{equation}
\lim_{t\to\infty}\mathcal{C}_2 (t; \vec{p}) = \frac{|Z_\pi(\vec{p})|^2}{2E_\pi(\vec{p})} \left[ e^{-E_\pi(\vec{p}) t} + e^{-E_\pi(\vec{p}) (T - t)} \right] .
\label{eq:2pt_corr}
\end{equation}

\section{Numerical Details}
\label{sec:numerical_details}
The configurations employed in this work are a superset of those employed in Ref.~\cite{Detmold:2021qln}. Details of the ensembles used are summarized in Tab.~\ref{tab:lattice_details}. The required 2- and 3-point functions were generated using the software packages \textsc{Chroma} with the \textsc{QPhiX} inverters \cite{chroma, qphix}, and 
the custom-built QC package~\cite{Grebe2025} with the QPhiX inverters~\cite{qphix}.

\begin{table*}
\begin{ruledtabular}
\begin{tabular}{ c c c c c c c c } 
$(L/a)^3 \times T/a$ & $\beta$ & $a$ (fm) & $\kappa_l$ & $\kappa_\text{H}$ & $c_\text{sw}$ & $N_\text{cfg}$ & $N_\text{meas}$ 
\\ \hline
$24^3 \times 48$ & ~6.10050~ & 0.0813 & ~0.134900~ & $\begin{matrix} 0.1300 \\ 0.1250 \\ 0.1200 \\ 0.1160 \\ 0.1100 \end{matrix}$ & ~1.6842~ & 6550 & 6550
\\ \hline
$32^3 \times 64$ & 6.30168 & 0.0600 & 0.135154 & $\begin{matrix} 0.1320 \\ 0.1280 \\ 0.1250 \\ 0.1184 \\ 0.1130 \\ 0.1095 \end{matrix}$ & 1.5792 & 7000 & 14000
\\ \hline
$40^3 \times 80$ & 6.43306 & 0.0502 & 0.135145 & $\begin{matrix} 0.1270 \\ 0.1217 \\ 0.1150 \end{matrix}$ & 1.5292 & 250 & 10800
\\ \hline
$48^3 \times 96$ & 6.59773 & 0.0407 & 0.135027 & $\begin{matrix} 0.1285 \\ 0.1244 \\ 0.1192 \\ 0.1150 
\end{matrix}$ & 1.4797 & 341 & 10000
\\
\end{tabular}
\end{ruledtabular}
\caption{Details of parameters used in this numerical study, where $L/a$ and $T/a$ are the spatial and temporal extents of the lattice in lattice units; $\beta=6/g^2$ is the inverse coupling; $a$ is the lattice spacing; $\kappa_l$ is the light quark hopping parameter, $\kappa_i^{-1}=2am_i+8$ (this study works in the isospin limit, where $m_u=m_d=m_l$); $\kappa_\text{H}$ is the heavy quark hopping parameter; $c_\text{sw}$ is the clover coefficient obtained from Ref.~\cite{Luscher:1996jn}; $N_\text{cfg}$ is the number of configurations employed; and $N_\text{meas}$ is the total number of measurements. In cases where this number is larger than the $N_\text{cfg}$, mmultiple source locations were used on each configuration.
}
\label{tab:lattice_details}
\end{table*}

\begin{figure}
\centering
\includegraphics[width=0.95\columnwidth]{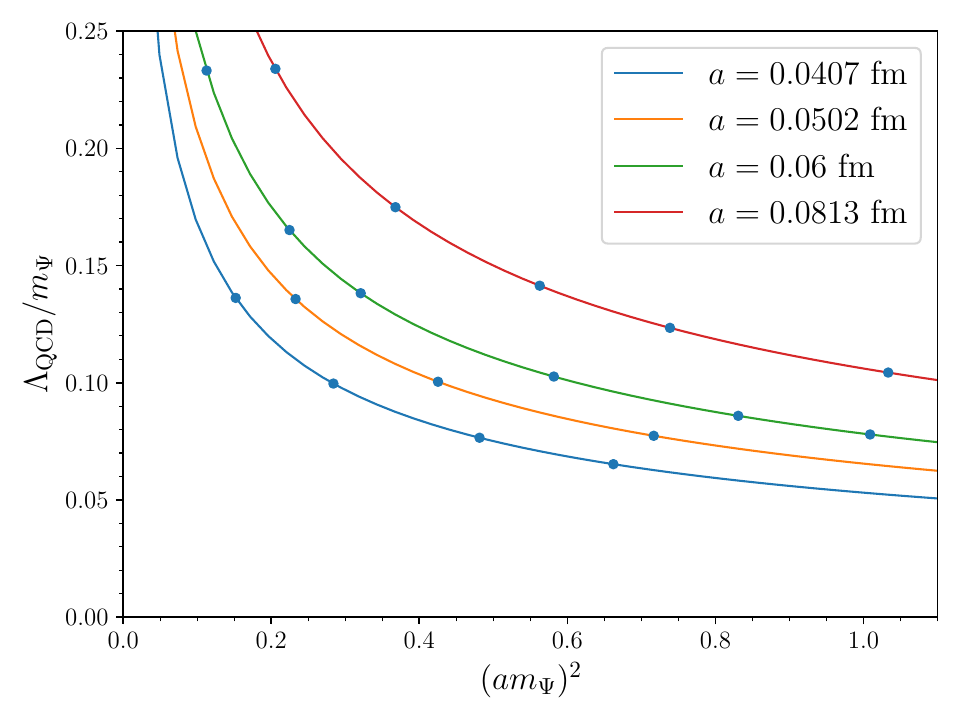}
\caption{Heavy quark masses considered in this work. The HOPE method assumes a hierachy of scales in which $\Lambda_\text{QCD}/m_\Psi \ll 1$ so that higher-twist contributions are suppressed. However, for $am_\Psi\sim 1$ large lattice artifacts are anticipated. Thus a variety of heavy quark masses are chosen to map out the higher-twist contributions and lattice artifacts so that they can both be removed.}
\label{fig:enter-label}
\end{figure}

\subsection{Overview of Analysis}
The extraction of Mellin moments of the pion LCDA from correlation functions computed using LQCD requires several steps of post-processing. To begin with, recall that LQCD provides direct access to the two- and three-point correlation functions given by Eqs.~\eqref{eq:2pt_corr} and \eqref{eq:3pt_corr}. From Eq.~\eqref{eq:3pt_corr}, it is clear that the large Euclidean time dependence of the 3-point correlation function  determines the hadronic matrix element of interest, $R^{\mu\nu}(t_-;\vec{p},\vec{q})$. In order to isolate this matrix element, the quantity
\begin{equation}
\begin{split}
\mathcal{R}^{\mu\nu}&(t_+,t_-;\vec{p},\vec{q})
=Z_A^{(0)2}(1+\tilde{b}_A a \tilde{m}_{ij})^2 
\\
&\times \bigg[\frac{Z_\pi(\vec{p})}{2E_\pi(\vec{p})}e^{-E_\pi(\vec{p})(t_e+t_m)/2}\bigg]^{-1} \times \mathcal{C}_3^{\mu\nu}(t_e, t_m; \vec{p}_e, \vec{p}_m)
\label{eq:data_ratio}
\end{split}
\end{equation}
is constructed. The factors $Z_\pi(\vec{p})$ and $E_\pi(\vec{p})$ are extracted from a standard spectroscopy analysis of the two-point correlation function given in Eq.~\eqref{eq:2pt_corr}. Details of this analysis are described in Sec.~\ref{sec:hadron_spectroscopy}, below. The multiplicative factor $Z_A^{(0)2}(1+ \tilde{b}_A a \tilde{m}_{ij})^2$ renormalizes the axial currents in the hadronic matrix element. $Z_A^{(0)}$ is the axial-vector operator renormalization constant calculated in the chiral limit, $\tilde{m}_{ij}=(m_i+m_j)/2$ is the average masses of the two quark fields in the heavy-light current and $\tilde{b}_A$ is a parameter which must be tuned to remove the $\mathcal{O}(a \tilde{m}_{ij})$ corrections to the axial-vector operator\footnote{This achieves full $\mathcal{O}(a)$ improvement of the operators used because, when evaluated in this matrix element, symmetries prevent the contributions from additional operators~\cite{Detmold:2021qln}.}. Values for these quantities are taken from Ref.~\cite{Luscher:1996jn} and Ref.~\cite{Bhattacharya:2005ss}, respectively. While the $Z_A^{(0)}$ factors are relatively well constrained, the $\tilde{b}_A$ factors are much less well known. In principle, an incorrect choice of $\tilde{b}_A$ will lead to the presence of $\mathcal{O}(a)$ lattice artifacts. However, since this amounts to a multiplicative mistuning, it will be wholly absorbed into $f_\pi$, which from Eq.~\eqref{eq:Gegen_OPE_had_amp} can be seen to serve as an overall normalization to the hadronic matrix element. Therefore, while the precise determination of $\tilde{b}_A$ is essential for an accurate extraction of $f_\pi$, it is irrelevant for the determination of the Mellin moments.

For suitably large values of $t_+$, the above ratio is expected to asymptote to the hadronic matrix element, $R^{\mu\nu}(t_-,\vec{p},\vec{q})$, that is,
\begin{equation}
\lim_{t_+\to\infty}\mathcal{R}^{\mu\nu}(t_+,t_-;\vec{p},\vec{q})=R^{\mu\nu}(t_-,\vec{p},\vec{q}).
\end{equation}
Corrections to this form at finite $t_+$ are exponentially suppressed by the mass-gap to the first excited state of the pion. By studying the $t_+$-dependence of $\mathcal{R}^{\mu\nu}$, it is possible to determine when residual excited state contamination is smaller than statistical uncertainties. For fixed $t_m$, $t_+\to\infty$ requires $t_e\to\infty$. Thus excited state contamination can be studied by examining the $t_e$ dependence of the hadronic matrix element. Rather than extrapolating data to $t_e\to\infty$, in this work $t_e$ is fixed in physical units to be large enough to suppress excited state contamination. This in turn means that for fixed $t_e$, larger $t_m$ have exponentially suppressed excited state contamination. Therefore it is particularly important to demonstrate that excited state contamination is suppressed at early Euclidean times. This point will be returned to later.

Without loss of generality, the hadronic matrix element can be decomposed into terms which are symmetric and antisymmetric in $t_-$: 
\begin{equation}
\mathcal{R}^{\mu\nu}(t_+,t_-;\vec{p},\vec{q})=\mathcal{R}_\text{sym}^{\mu\nu}(t_+,t_-;\vec{p},\vec{q})+\mathcal{R}_\text{anti}^{\mu\nu}(t_+,t_-;\vec{p},\vec{q})
\end{equation}
where
\begin{align}
\label{eq:sym}
\begin{split}
\mathcal{R}_\text{sym}^{\mu\nu}&(t_+,t_-;\vec{p},\vec{q})
\\
&=\frac{1}{2}\bigg[\mathcal{R}^{\mu\nu}(t_+,t_-;\vec{p},\vec{q})+\mathcal{R}^{\mu\nu}(t_+,-t_-;\vec{p},\vec{q})\bigg],
\end{split}
\\
\label{eq:antisym}
\begin{split}
\mathcal{R}_\text{anti}^{\mu\nu}&(t_+,t_-;\vec{p},\vec{q})
\\
&=\frac{1}{2}\bigg[\mathcal{R}^{\mu\nu}(t_+,t_-;\vec{p},\vec{q})-\mathcal{R}^{\mu\nu}(t_+,-t_-;\vec{p},\vec{q})\bigg].
\end{split}
\end{align}
The advantage of this decomposition is threefold.  First, as shown previously, $\gamma_5$-hermiticity can be used to relate the hadronic matrix elements at positive and negative $t_-$ values to the hadronic matrix elements with different three-momenta~\cite{Detmold:2021qln}. Thus it can be shown that both the even and odd pieces can be determined from a study of correlation functions for $t_->0$. This enables the first current operator to be placed closer to the source, reducing the statistical uncertainties stemming from Euclidean time evolution. In addition, it has been shown in Refs.~\cite{Detmold:2020lev,Detmold:2021qln} that the computation of quantities using only positive $t_-$ data leads to more correlated statistical fluctuations which cancel when the required linear combinations are taken. 

\begin{figure}
    \centering
    \includegraphics[scale=0.5]{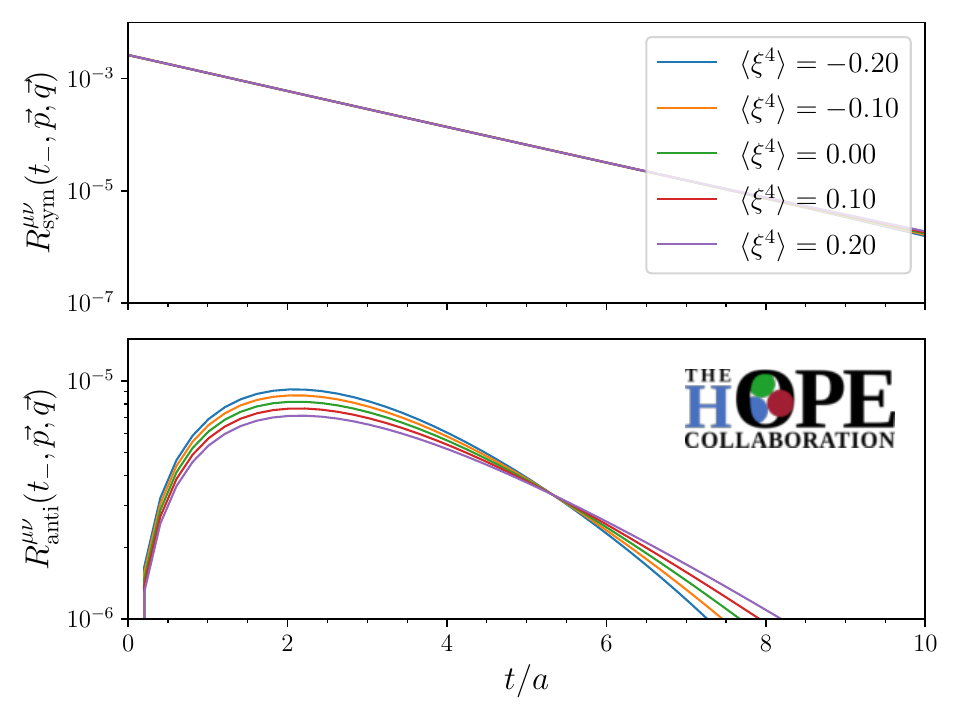}
    \caption{Visualizing the predictions of the Fourier transform of the one-loop HOPE formula given in Eq.~\eqref{eq:Gegen_OPE_had_amp} for special kinematics where the leading term in $R_\text{anti}^{\mu\nu}(t_-,\vec{p},\vec{q})$ is due to $\braket{\xi^2}$. Therefore, $\braket{\xi^4}$ is a small but has significant contribution to $R_\text{anti}^{\mu\nu}(t_-,\vec{p},\vec{q})$, while it has is negligible contribution to $R_\text{sym}^{\mu\nu}(t_-,\vec{p},\vec{q})$. This is demonstrated here for the kinematics desibed in the text for $L/a = 32$, $a E_\pi = 0.4271$ $a f_\pi = 0.05$, $am_\Psi = 0.7$ and $\braket{\xi^2} = 0.2$. }
    \label{fig:hope_kinematics}
\end{figure}

Finally, it has been demonstrated in Refs.~\cite{Detmold:2020lev,Detmold:2021qln} that using 
special kinematics where $p_3=0$ and $\mu,\nu=1,2$ results in the leading non-zero contribution to $\mathcal{R}_\text{anti}^{\mu\nu}(t_-,\vec{p},\vec{q})$ being due to $\braket{\xi^2}$, and thus the first correction to this is due to $\braket{\xi^4}$. This is displayed in Fig.~\ref{fig:hope_kinematics}.
For this study, the momenta
\begin{align}
\vec{p}&=\frac{2\pi}{L}(2,0,0),
\\
\vec{q}&=\frac{2\pi}{L}(1,0,-1)
\end{align}
are employed. The statistical signal is increased by averaging the correlation function over equivalent momenta.

Having constructed the symmetric and anti-symmetric parts of the hadronic tensor $\mathcal{R}^{\mu\nu}(t_e,t_m;\vec{p},\vec{q})$, a fit to the continuum, twist-2 form of the HOPE given in Eq.~\eqref{eq:Gegen_OPE_had_amp} with Wilson coefficients computed to one-loop is performed. The effect of fitting data at finite lattice spacing with the continuum, twist-2 form of the HOPE formula implies that the extracted parameters differ from their continuum, twist-2 values by lattice artifacts and higher-twist contributions. These contributions are removed by fitting the extracted parameters to a model which incorporates the both the leading lattice artifacts and higher-twist contributions. In the following subsections, each of these computational steps is described in more detail.

\subsection{Hadron spectroscopy}\label{sec:hadron_spectroscopy}
In order to isolate the ground-state energy, $E_\pi(\vec{p})$, and corresponding overlap factor, $Z_\pi(\vec{p})$, standard techniques from hadron spectroscopy are employed. In particular, the variational method~\cite{Michael:1982gb,Luscher:1990ck} is employed to produce an optimized pseudoscalar interpolating operator, which is consistently used in the construction of the two- and three-point correlation functions studied in this work. 

\begin{figure*}
\centering
\includegraphics[width=0.45\linewidth]{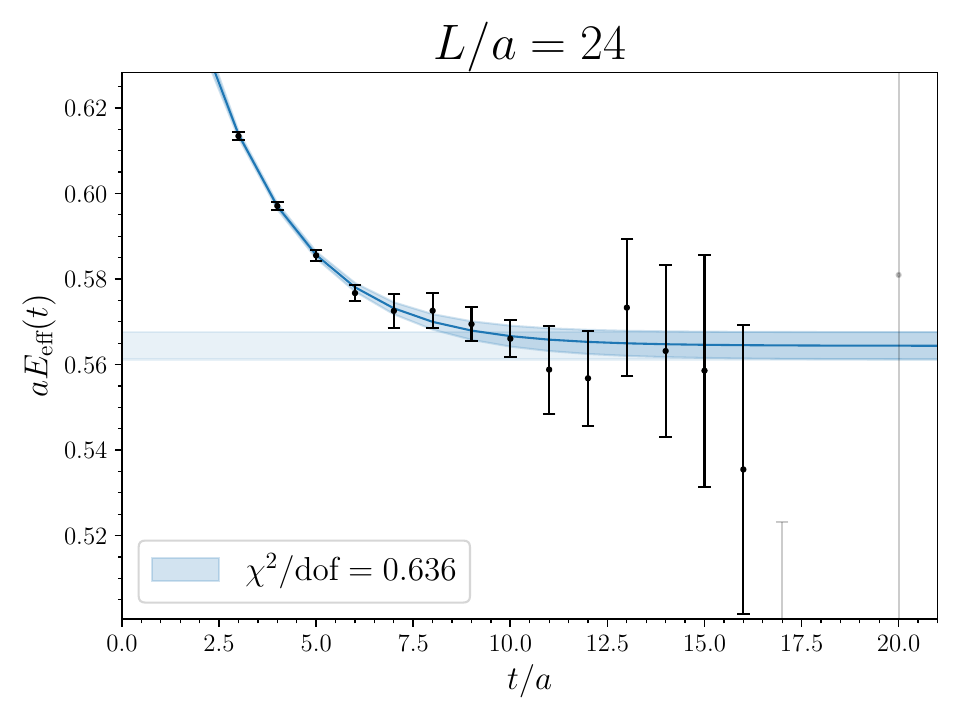}
\includegraphics[width=0.45\linewidth]{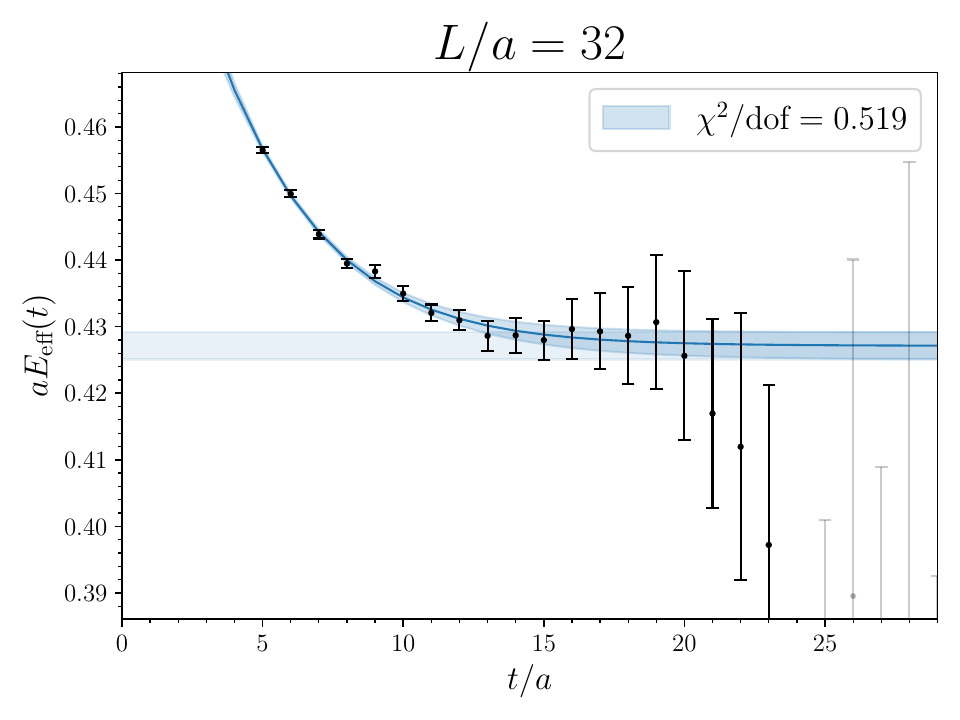}

\includegraphics[width=0.45\linewidth]{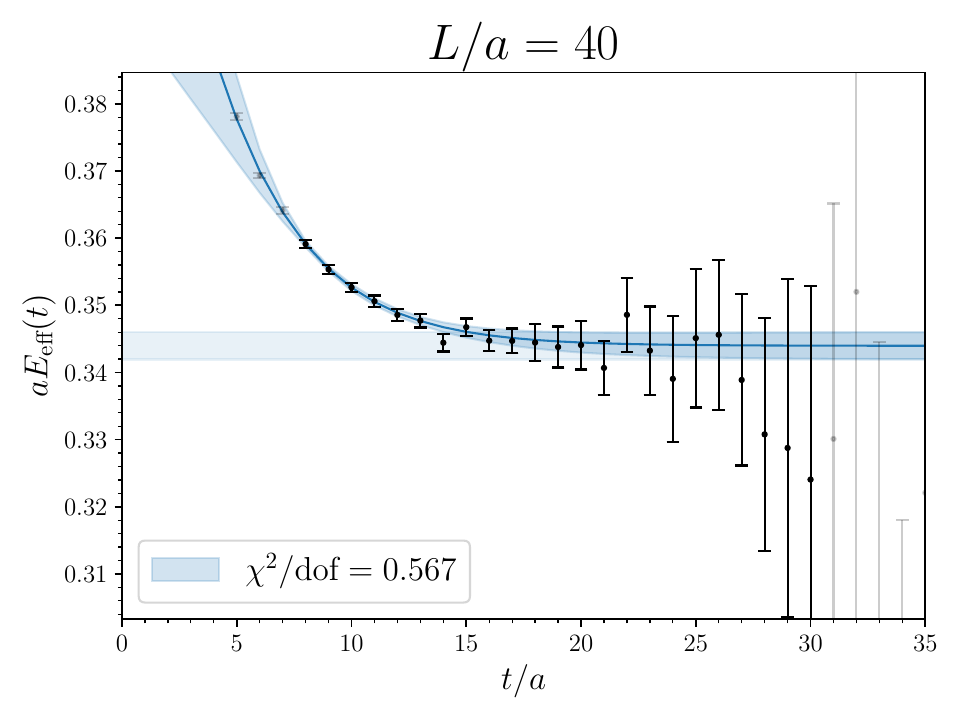}
\includegraphics[width=0.45\linewidth]{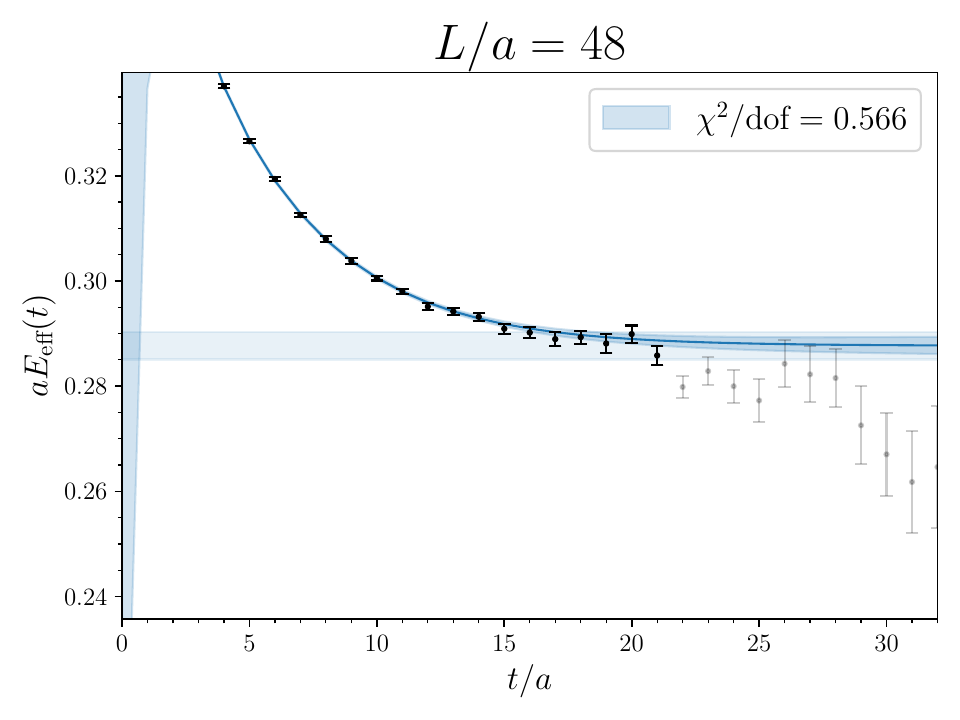}
\caption{Effective mass plots for ensembles studied in this work. Grey data points denote the numerical data not used in the fit, black points indicate that they are included in the fit, blue curves show the highest weight fit, and the shaded horizontal bands denote the resulting ground state masses extracted from this analysis.
}
\label{fig:effective_mass}
\end{figure*}

The starting point for the variational method is to propose a set of operators with the quantum numbers of the state one is interested in studying. In this work, the operator set is defined by $\mathbb{S}_\pi=\{\mathcal{O}_1(x),\mathcal{O}_2(x)\}$,
where
\begin{align}
\mathcal{O}_1(x)&=\overline{\psi}(x)\gamma_5\psi(x), 
\\
\mathcal{O}_2(x)&=\overline{\psi}(x)\gamma_4\gamma_5\psi(x).
\end{align}
$\mathcal{O}_2(x)$ was recently found to enhance the signal-to-noise ratio of the two-point and three-point pion correlation functions by a factor that is proportional to the square of the Lorentz boost factor~\cite{Zhang:2025hyo}.
The quark fields, $\psi(x)$ are spatially smeared using momentum smearing~\cite{Bali:2016lva} accelerated using the Frigo-Strumpen algorithm~\cite{Frigo2007-ss} to increase their overlap with boosted hadron states. With these two operators, a $2\times2$ matrix of correlation functions can be constructed:
\begin{equation}
\mathcal{C}_{ij}(t,\vec{p})=\sum_{\vec{x}\in |\Lambda|} e^{i\vec{p}\cdot \vec{x}}\braket{0|\mathcal{O}_i(x)\mathcal{O}_j^\dagger(0)|0}.
\end{equation}
This matrix of correlation functions can be diagonalized through a generalized eigenvalue problem (GEVP),
\begin{equation}
\sum_{j}\mathcal{C}_{ij}(t,\vec{p})v_{j,n}(t,t_0)=\sum_{j}\lambda_n(t-t_0)\mathcal{C}_{ij}(t_0,\vec{p}) v_{j,n}(t,t_0),
\end{equation}
where $v_{j,n}(t,t_0)$ is the $j$th component of the $n$th eigenvector that corresponds to the $n$th eigenvalue $\lambda_{n}$, and $n \in \{0, 1\}$. Without loss of generality, the eigenvalues are ordered so that $\lambda_n(t-t_0)\geq\lambda_m(t-t_0)$ for $n<m$. It is possible to construct an optimized interpolating operator for a particular state in the spectrum by employing 
\begin{equation}
\label{eq:optimized_interpolating_operator}
\mathcal{O}_\pi(x)=\sum_{i=1}^2 v_{0}^{i}(t_\text{ref},t_0) \mathcal{O}_i(x),
\end{equation}
where $t_0/a=\{6,8,10,12\}$ and $t_\text{ref}/a=\{12,16,20,24\}$ for $L/a=\{24,32,40,48\}$.
The diagonal elements of the matrix
\begin{equation}
\tilde{\mathcal{C}}_{mn}(t,\vec{p})=\sum_{i,j=1}^2 v_{m}^{i} \mathcal{C}_{ij}(t,\vec{p})v_{n}^{j},
\end{equation}
are positive-definite sums of decaying exponentials. Thus the effective mass of $\tilde{\mathcal{C}}_{00}(t,\vec{p})$ 
\begin{equation}
aE_\text{eff}(t)=\ln(\frac{\tilde{\mathcal{C}}_{00}(t,\vec{p})}{\tilde{\mathcal{C}}_{00}(t+1,\vec{p})})
\end{equation}
provides a stochastic variational upper bound on the ground state of the system, that is, $E_\text{eff}(t)\geq E_\pi(\vec{p})$.\footnote{In this work, it will be assumed that the variational bound is saturated at large Euclidean times, so that a fit to this region gives information on the underlying eigenvalues of the QCD transfer matrix.}
In practice, the correlation function $\tilde{C}_{00}(t,\vec{p})$ is fit to a truncated sum of exponentials of the form
\begin{equation}
{C}(t)=\sum_{s=0}^{N_\text{state}} \frac{|Z_s|^2}{2E_s} e^{-E_s t},
\end{equation}
where $N_\text{state}$ is the number of energy levels included in the model. In order to constrain the energies $E_s$ and overlap factors $|Z_s|^2$, data in the range $[t_\text{start},t_\text{stop}]$  is fit to a model with $N_\text{state}$ energies. The value of $t_\text{stop}$ is taken as the largest time-slice where the data satisfies 
\begin{equation}
\frac{\sigma_{\tilde{\mathcal{C}}_{00}}(t)}{\tilde{\mathcal{C}}_{00}(t)}<0.1, 
\end{equation}
where $\sigma_{\tilde{\mathcal{C}}_{00}}(t)$ is the statistical uncertainty in the correlator at time $t/a$, 
while the optimal $t_\text{start}$ and $N_\text{state}$ are determined by maximizing 
\begin{equation}
\label{eq:model_weights}
w\propto \exp(-\text{AIC}/2-N_\text{cut}),
\end{equation}
where $\text{AIC}=\chi^2+2k$ is the Akaike information criterion~\cite{Akaike1974}, $k$ is the number of parameters included in the fit, and $N_\text{cut}$ is the number of timeslices removed from the fit. This term therefore serves to penalize models which drop early time data points. This cost function is motivated by Ref.~\cite{Jay:2020jkz} where these weights are employed to perform model averaging. The $\chi^2$-function is 
\begin{equation}
\chi^2=\sum_{t,t^\prime} (\tilde{\mathcal{C}}_{00}(t,\vec{p})-{C}(t))[\Sigma^{-1}]_{t,t^\prime}(\tilde{\mathcal{C}}_{00}(t^\prime,\vec{p})-{C}(t^\prime)),
\end{equation}
where $\Sigma_{t,t^\prime}$ is the covariance of the data. The resulting fits for the four ensembles considered in this work are shown in Fig.~\ref{fig:effective_mass}. The fitted ground-state energies are given in Tab.~\ref{tab:two-point_energies}.

\begin{table}
\begin{ruledtabular}
\centering
\begin{tabular}{c c c c}
$L/a$ & $L$ (fm)  & $aE_0(\vec{p})$ & $E_0(\vec{p})$ (GeV) \\ \hline
24 & 1.95 & 0.5644(32)& 1.3676(77) \\
32 & 1.92 & 0.4271(21)& 1.4024(67) \\
40 & 2.01 & 0.3445(14)& 1.3520(54) \\
48 & 1.95 & 0.2876(25)& 1.392(13) \\
\end{tabular}
\end{ruledtabular}
\caption{Ground state energy for the pion with momentum $|\vec{p}| L/(2\pi) = 2$.}
\label{tab:two-point_energies}
\end{table}

\subsection{Studying the hadronic matrix element}
The correlator defined by Eq.~\eqref{eq:3pt_corr} is computed using the  pion interpolating operator defined in  Eq.~\eqref{eq:optimized_interpolating_operator}.
This leads to $C_i^{\mu\nu}(t_e, t_m; \vec{p}_e, \vec{p}_m),~i=1,2$. Employing the optimized weights obtained from the variational analysis described above, the matrix element for the optimized pseudoscalar interpolating operator, $\mathcal{O}_\pi(x)$, is constructed via
\begin{equation}
\mathcal{C}_3^{\mu\nu}(t_e, t_m; \vec{p}_e, \vec{p}_m)=\sum_{i=1}^2 v_{i,0} \mathcal{C}_i^{\mu\nu}(t_e, t_m; \vec{p}_e, \vec{p}_m).
\end{equation}
Using the best fit values for the ground-state energy $E_\pi(\vec{p})$ and corresponding overlap factor $Z_\pi(\vec{p})$, the ratio defined in Eq.~\eqref{eq:data_ratio} is constructed.

\begin{figure}
\centering
\includegraphics[scale=0.5]{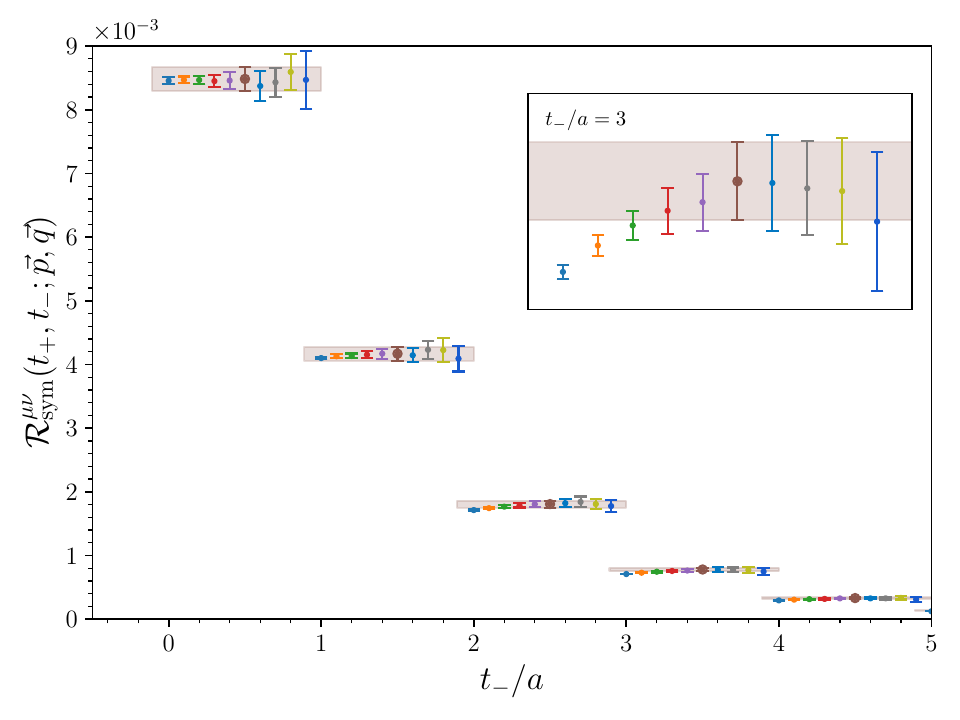}
\includegraphics[scale=0.5]{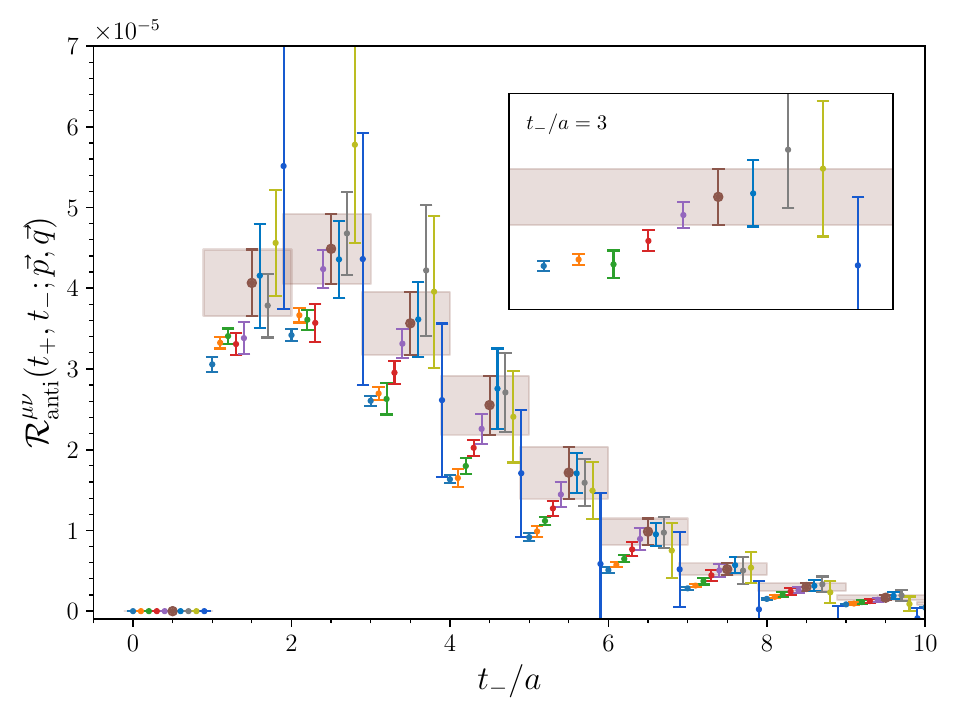}
\caption{Excited state contamination in even and odd components of ratio $\mathcal{R}^{\mu\nu}(t_e,t_m;\vec{p},\vec{q})$ for a sequential source location $t_e/a\in \{4,5,6,\dots,12\}$. Results are shown for the $L/a=24$ ensemble with $N_\text{meas}=2000$ and
$\kappa_\text{H}=0.120$. 
Excited state contamination is negligible compared to statistical uncertainties for $t_e/a\geq8$ in all cases. The inset plot presents a zoomed in version of $t_-/a=3$ showing the $t_e/a$ dependence. The shaded band extends the statistical errors from the $t_e/a=8$ point to demonstrate the convergence of these results within statistical uncertainties.}
\label{fig:ratio_excited_states}
\end{figure}

Excited state contamination is assessed by studying the dependence on the ratios $\mathcal{R}_\text{sym}^{\mu\nu}(t_+,t_-,\vec{p},\vec{q})$ and $\mathcal{R}_\text{anti}^{\mu\nu}(t_+,t_-,\vec{p},\vec{q})$
as a function of $t_e$. States which propagate between the source and current must possess the quantum numbers of the pion, and due to quenching cannot contain heavy sea quarks.  Thus, it is expected that excited state contamination should be approximately independent of the heavy-quark mass.  This contamination is studied using the $L/a=24$ ensemble and the resulting ratio $\mathcal{R}^{\mu\nu}(t_+,t_-;\vec{p},\vec{q})$ is shown in Fig.~\ref{fig:ratio_excited_states}. From this, it can be seen that statistical uncertainties dominate excited state contamination for $t_e/a=8$. This corresponds approximately to $t_e/a=\{11,13,16\}$ for $L/a=\{32,40,48\}$. In this work, data at these fixed $t_e/a$ are used
with the assumption that excited state effects are similar in each ensemble at comparable physical time separations. 

\begin{figure}
\centering
\includegraphics[width=\linewidth]{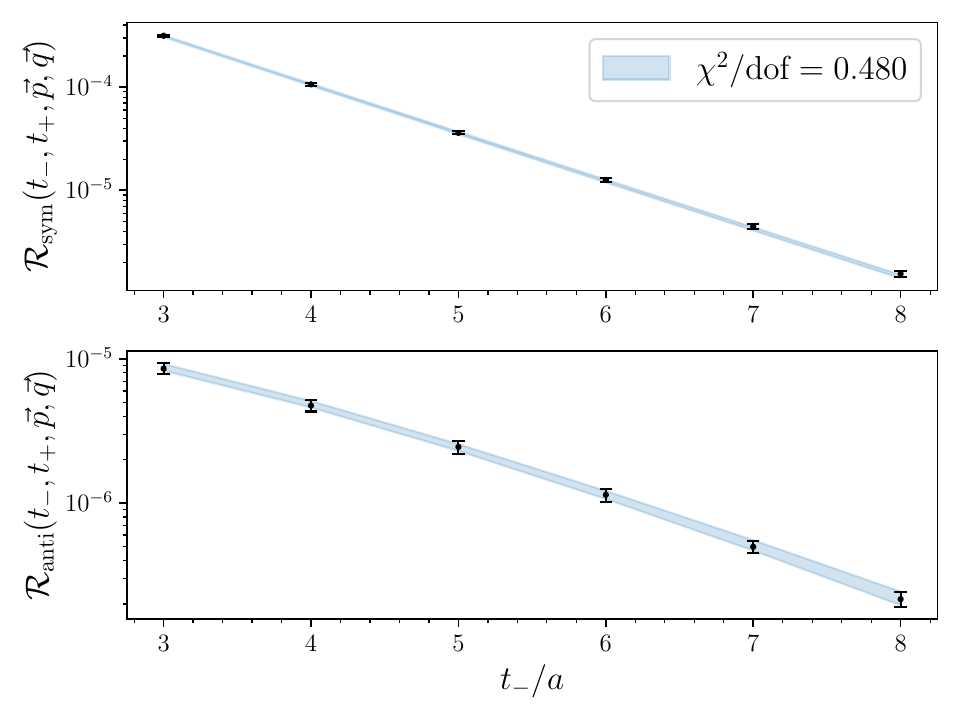}
\caption{Example fit to the time-momentum equivalent of Eq.~\eqref{eq:Gegen_OPE_had_amp}, for $L/a=24$ data with $\kappa_\text{H}=0.11$. The shaded band represents the one-sigma uncertainties in fitted HOPE model. Fits with highest weight for all ensembles are shown in Appendix~\ref{app:hope_fits}. 
}
\label{fig:hope_fit}
\end{figure}

In order to learn about the Mellin moments of the pion LCDA and higher twist corrections, it is necessary to fit the numerical ratio $\mathcal{R}^{\mu\nu}(t_+,t_-;\vec{p},\vec{q})$ computed at a range of heavy quark masses to the HOPE formula given by Eq.~\eqref{eq:Gegen_OPE_had_amp}, Fourier transformed to the time-momentum representation. An example of the resulting fitted ratio is shown in Fig.~\ref{fig:hope_fit}. 

In this work, the HOPE formula is evaluated at a fixed renormalization scale of $\mu = 2~\si{GeV}$. Other choices for this scale, including using a renormalization scale which varies with $\tilde{Q}^2$ or $t_-$ (for example) combined with renormalisation-group techniques for resumming possible large logarithms, are beyond the scope of this work.  Such techniques will become essential once more precise data are available in the future. 
 
Since lattice data is fit to the twist-2 continuum form of the HOPE formula, it is natural to expect that discrepancies between the HOPE model and the lattice data should occur at short distances, where residual excited states contamination may be present in the ratio, $\mathcal{R}^{\mu\nu}(t_-,t_+,\vec{p},\vec{q})$, and lattice artifacts are not suppressed~\cite{DellaMorte:2008xb,Ce:2021xgd}, and also at large distances, where higher-twist contributions and nonperturbative effects can be significant. Therefore, there exists some uncertainty about the exact range of Euclidean time which should be included in the fit. This poses a challenge for the analysis, because the fitted parameters implicitly depend on the lattice spacing and heavy-quark mass, that is, $\braket{\xi^n}=\braket{\xi^n}(a,m_\psi)$. Therefore, the final step in this analysis requires a continuum, twist-2 extrapolation. Empirically, it was found that using a single fit window in $t_{-}$ in the HOPE-formula fits for all heavy quark masses and lattice spacings led to unacceptably large $\chi^2/\text{dof}$ when subsequent continuum, twist-2 extrapolations were attempted with the full dataset, despite the fact that individual $\chi^2/\text{dof}$ values for each fit to the HOPE data were reasonable. 

Intuitively, this is because the statistical errors from individual fits of the HOPE formula to the numerical ratio $\mathcal{R}^{\mu\nu}(t_+,t_-,\vec{p},\vec{q})$ underestimate the systematic uncertainties in $\braket{\xi^2}$ and $\braket{\xi^4}$. A natural framework for dealing with this problem is provided by model averaging~\cite{Jay:2020jkz}. While this prescription is attractive because it provides a simple, theoretically well-motivated approach to incorporating the variance over models into the total uncertainty, it is difficult to implement in a multi-fit bootstrap analysis like this one, where the results of this fit are used as inputs into a second set of fits\footnote{The challenge of combining bootstrap techniques and model averaging was acknowledged as an outstanding challenge in Ref.~\cite{Jay:2020jkz}.}. One solution to this problem is to interpret the series of analysis choices, including the $t_\text{start}$ and $t_\text{stop}$ \textit{as well as} the subsequent twist-2, continuum extrapolation as a single model and to perform a model average over this set of ``meta-models.'' In practice, this approach leads to a large number of models, which limit the feasibility of this strategy. Instead, in this work the following heuristic method is taken. 

In order to determine the region where the HOPE model is appropriate, fits are performed over a range of $t_\text{start}/a\in[3,10]$ and $t_\text{stop}/a=(t_\text{start}+\Delta)/a$ with $\Delta/a\in[5,10]$. Each fit is assigned a weight again, according to Eq.~\eqref{eq:model_weights}. Rather than using a single combination of $(t_\text{start},\Delta)$ which optimizes the weight, the choice of $(t_\text{start},\Delta)$ is varied bootstrap by bootstrap, with random selection implemented according to the above weights such that fit windows with large probability are sampled more often than fits with small probability. In this way the variance in fitted parameters over the space of $(t_\text{start},\Delta)$ are incorporated into the bootstrap samples. Example fits for all ensembles are shown in Appendix~\ref{app:hope_fits}. As a result, the width of the distribution of bootstrap means no longer has a rigorous relationship to the statistical uncertainty and instead should be thought of as empirically encoding the combined statistical and systematic variance stemming from the choice of $(t_\text{start},\Delta)$. 

The results of these fits are a set of bootstrapped values for $\{m_\Psi, f_\pi,\langle \xi^2\rangle,\langle \xi^4\rangle\}$. Since these parameters are obtained from fitting the continuum, twist-2 HOPE formula to lattice data, the resulting parameters contain lattice artifacts and contributions from higher-twist effects. In the following section the extrapolation to the continuum, twist-2 limit is described. 

\subsection{Continuum, Twist-2 Extrapolation}

\begin{figure*}
\centering
\includegraphics[width=0.45\linewidth]{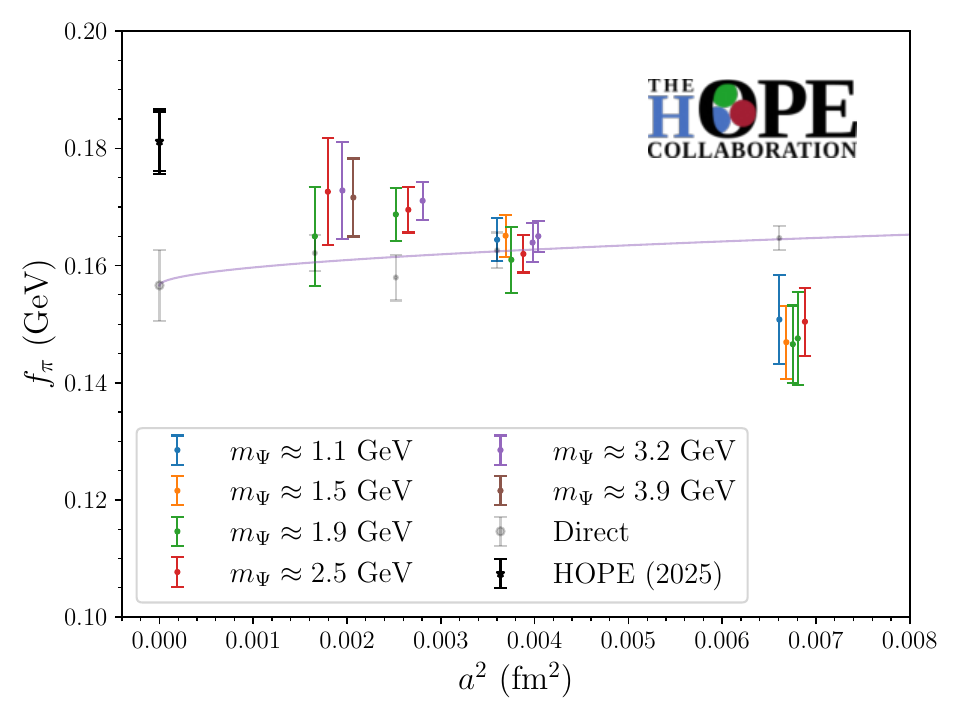}
\includegraphics[width=0.45\linewidth]{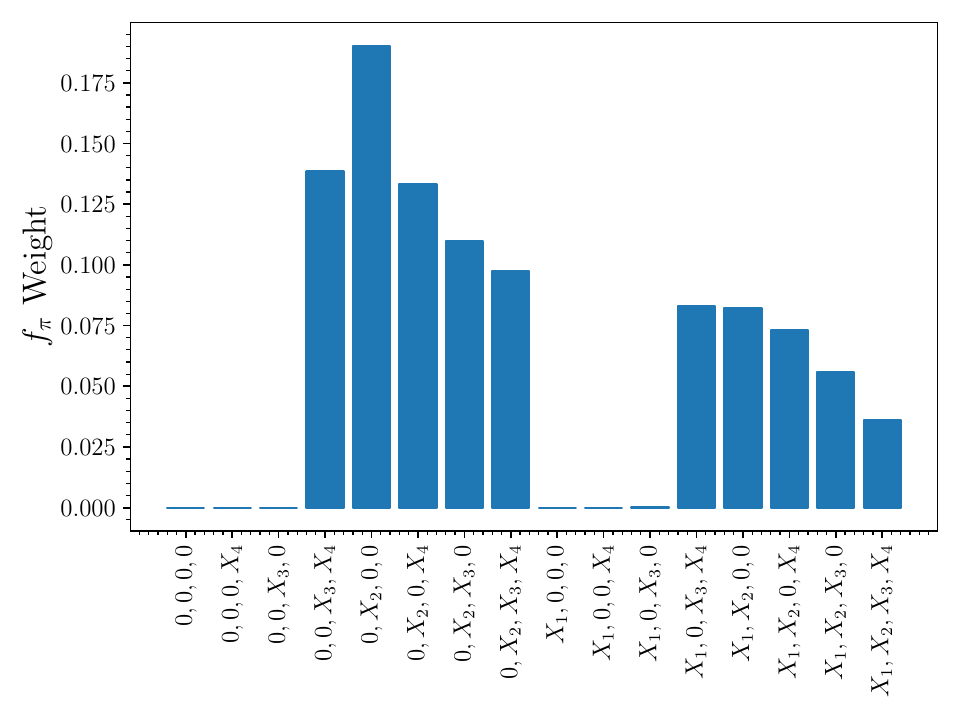}
\caption{Continuum, twist-2 extrapolation of $f_\pi$. The left-hand plot shows values of $f_\pi$ determined at different heavy quark masses, with the resulting extrapolated value in black. Data at finite lattice spacing are colored by the ensemble and offset proportional to the heavy-quark mass. Shaded grey data and curve correspond to two-point correlator analysis on a subset of the configurations studied here. 
The right-hand plot shows the relative weights, $w_i$ for the models considered in the model average. 
}
\label{fig:fpi}
\end{figure*}

\begin{figure*}
\centering
\includegraphics[width=0.45\linewidth]{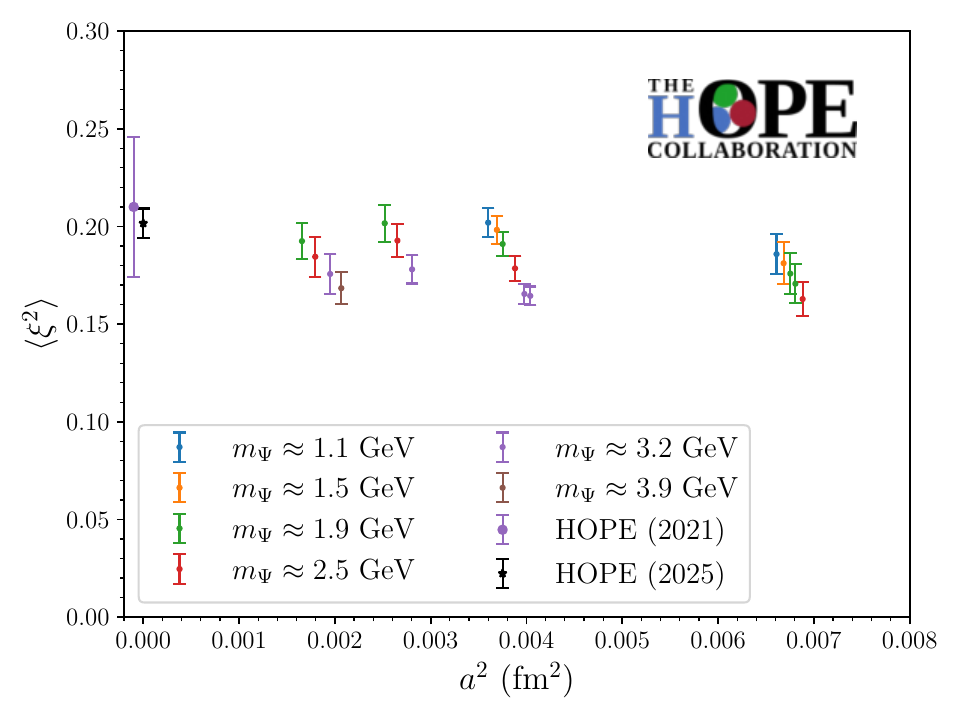}
\includegraphics[width=0.45\linewidth]{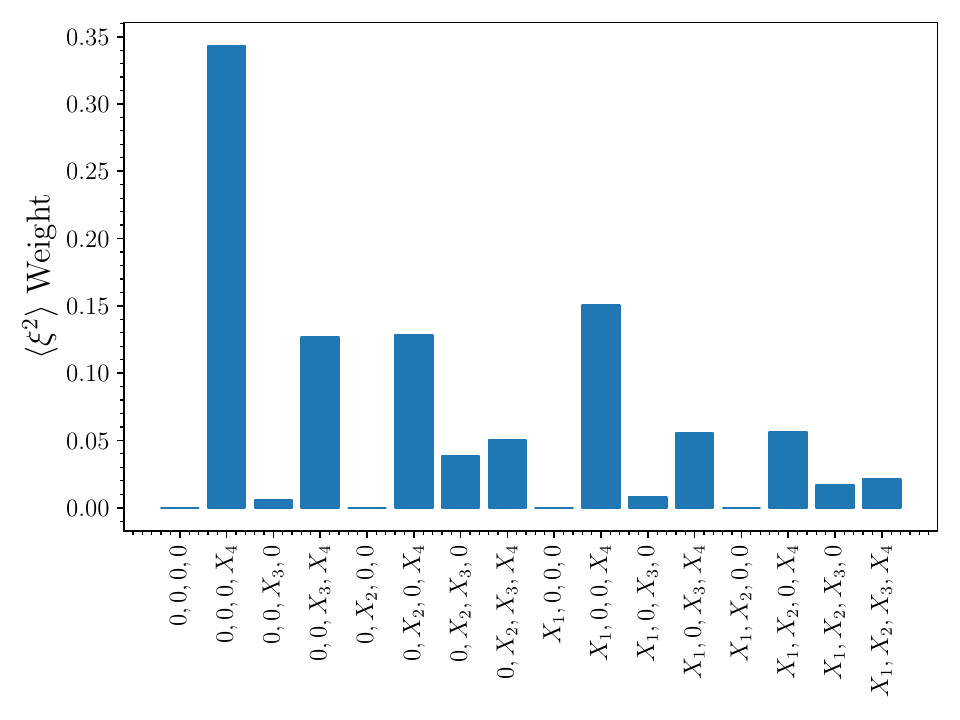}
\caption{Continuum, twist-2 extrapolation of second Mellin moment of pion LCDA. Details are as in Fig.~\ref{fig:fpi}.}
\label{fig:ctm_twist2_xi2}
\end{figure*}

\begin{figure*}
\centering
\includegraphics[width=0.45\linewidth]{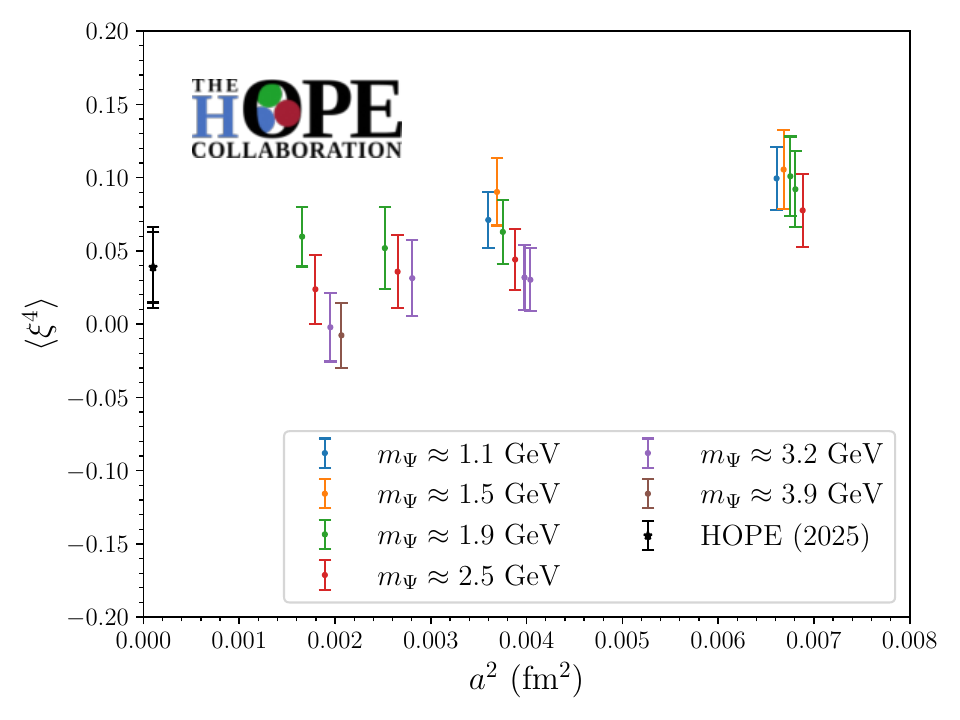}
\includegraphics[width=0.45\linewidth]{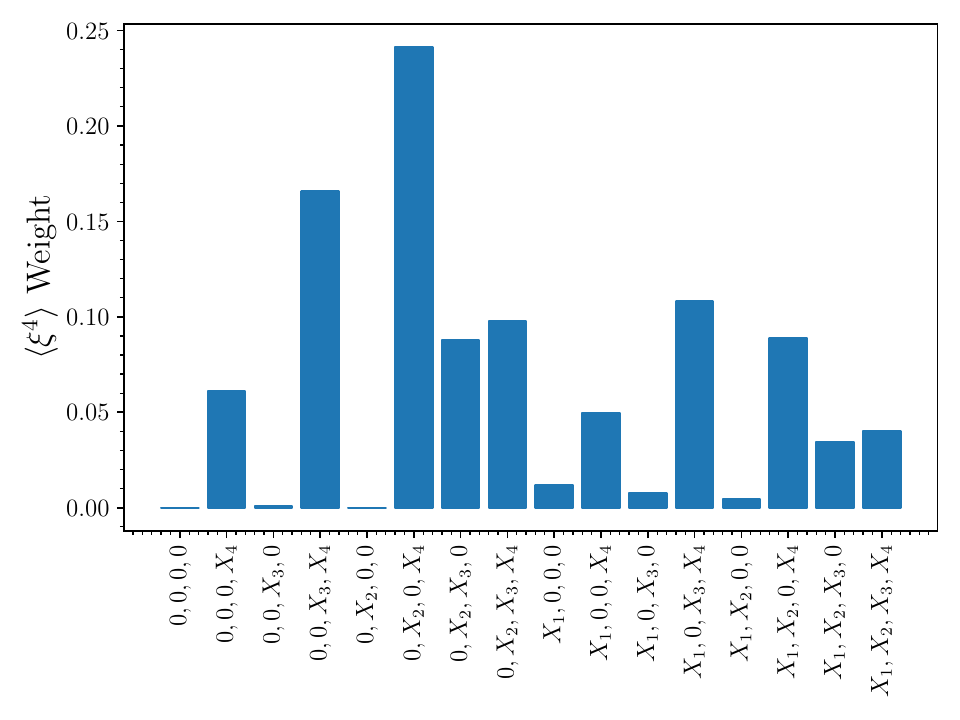}
\caption{Continuum, twist-2 extrapolation of fourth Mellin moment of pion LCDA. Continuum, twist-2 extrapolation of second Mellin moment of pion LCDA. Details are as in Fig.~\ref{fig:fpi}.
}
\label{fig:ctm_twist2_xi4}
\end{figure*}

Lattice and higher-twist 
artifacts are removed by fitting the data to a model of the form
\begin{equation}
\begin{split}
  X(a,m_\Psi) &= X_0 + \frac{X_1}{m_\Psi} + X_2 a^2 
  \\
  &+ X_3 a^2 m_\Psi + X_4 a^2 m_\Psi^2 \, ,
\end{split}
  \label{eq:ctm_twist2_model}
\end{equation}
where 
$X\in\{f_\pi,\braket{\xi^2},\braket{\xi^4}\}$
, $X_0$ is the parameter of interest, $X_1$, $X_2$, $X_3$, and $X_4$ are nuisance fit parameters that contain information regarding the sizes of the relevant artifacts. In order to provide an estimate of the systematic uncertainties stemming from the use of this procedure, model averaging~\cite{Jay:2020jkz} is employed. In particular, data shown in each of Figs.~\ref{fig:fpi},~\ref{fig:ctm_twist2_xi2},~\ref{fig:ctm_twist2_xi4} is fit to Eq.~\eqref{eq:ctm_twist2_model}, and all nested models obtained from setting all possible subsets of $\{X_1,X_2,X_3,X_4\}$ to zero. Following the prescription in Ref.~\cite{Jay:2020jkz}, for each model a weight $w_i\propto\exp(-\text{AIC}/2)$ is assigned. Note that since no data are removed from the fit, $N_\text{cut}=0$. The mean and variance are computed using
\begin{align}
\label{eq:jay}
\braket{X_j}&= \sum_i w_i \braket{X_j}_i,
\\
\sigma_{X_j}^2&=\sum_i w_i \sigma_{{X_j},i}^2 + \sum_i w_i\braket{X_j}_i^2 - \bigg[\sum_i w_i\braket{X_j}_i\bigg]^2.
\end{align}
Reference~\cite{Jay:2020jkz} suggests to interpret the first term in Eq.~\eqref{eq:jay} as the uncertainty due to statistical variation, while the second and third terms can be understood as the variance from model variation and therefore can be thought of as the systematic uncertainty.

In principle, the model described in Eq.~\eqref{eq:ctm_twist2_model} can be applied to $f_\pi$, $\braket{\xi^2}$ and $\braket{\xi^4}$. However, in practice,  $\braket{\xi^2}$ and $\braket{\xi^4}$ are independent of the absolute normalization of the hadronic matrix element $R^{\mu\nu}(t_-,\vec{p},\vec{q})$, whereas $f_\pi$ depends on this normalization. As discussed previously, this absolute normalization depends on the axial-vector renormalization $Z_A^{(0)}$ and $\tilde{b}_A$. Since $f_\pi$ is not the focus of this work, systematic errors from the estimation of these parameters are not propagated into the hadronic matrix element, and subsequently, a thorough estimation of uncertainties of $f_{\pi}$ is not performed.  In other words, several systematic errors on this quantity in the twist-2 continuum limit are not shown in the left plot of Fig.~\ref{fig:fpi} Given this situation, the agreement between the HOPE method and the conventional strategy of analysing two-point functions serves as a validation of the HOPE approach presented here.

Results for the second and fourth Mellin moments are shown in Figs.~\ref{fig:ctm_twist2_xi2} and \ref{fig:ctm_twist2_xi4}, respectively. The final determinations of the second and fourth Mellin moments are
\begin{align}
\braket{\xi^2}(\mu = 2~\si{GeV}) &= 0.202(8),
\\
\braket{\xi^4}(\mu = 2~\si{GeV}) &= 0.039(28).
\end{align}
Importantly, this value of the second Mellin moment is consistent with the previous determination~\cite{Detmold:2021qln} which utilized a subset of the configurations considered here and a smaller pion momentum.  

In order to estimate the uncertainty arising from the use of one-loop perturbation theory for the Wilson coefficents, the full analysis method is performed at a renormalization scale of $\mu^{\prime 2} = 2 \mu^2$. The result of this procedure gives the central values $\braket{\xi^2}(\mu = 2.8~\si{GeV}) = 0.211$, $\braket{\xi^4}(\mu = 2.8~\si{GeV}) = 0.054$. Evolving these moments to $\mu = 2~\si{GeV}$ using one-loop running leads to
$\braket{\xi^2}(\mu = 2~\si{GeV}) = 0.212$ and $\braket{\xi^4}(\mu = 2~\si{GeV}) = 0.050$. The difference between the central values of these two determinations at $\mu = 2~\si{GeV}$ is taken as a systematic error, to be added in quadrature with the combined statistical and systematic error from the variation over models. Therefore, the final results are 
\begin{align}
\braket{\xi^2}(\mu = 2~\si{GeV}) &= 0.202(8)(9),
\\
\braket{\xi^4}(\mu = 2~\si{GeV}) &= 0.039(28)(11).
\end{align}
In principle, the estimate of scale variation uncertainty should be performed by re-summing the logarithms in the $t_-$-space first, similar to those done in Refs.~\cite{Gao:2021hxl,Su:2022fiu,Baker:2024zcd,Cloet:2024vbv}. By Fourier transforming the logarithm of $\mu^2/(\tilde Q^2 + m_\psi^2$) in the momentum-space Wilson coefficient, one finds $\ln \big[\mu^2t_-/(2\sqrt{\vec{q}^2+m_\psi^2})\big]$ in $t_-$-space, which corresponds to the initial scale $[\mu_0(t_-)]^2= 2\sqrt{\vec{q}^2+m_\psi^2}/t_-$. Within the range of $t_-$ used for the fit, this scale remains above $\sim 1.5$ GeV, which is not much different from $\mu=2$ GeV and should not constitute a significant scale variation. However, the full OPE expression includes more complicated logarithms after the Fourier transform, which corresponds to different initial scales. A further study of this resummation will be carried out in a future work.

\section{Discussion}\label{sec:discussion}
\begin{figure*}
\centering
\includegraphics[width=0.8\linewidth]{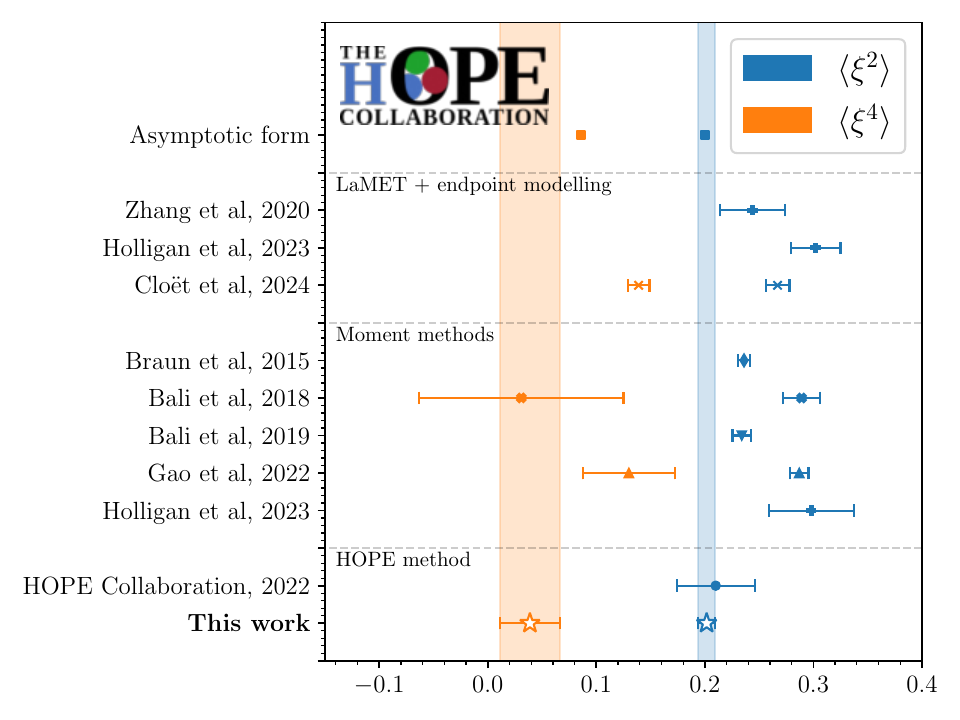}
\caption{Comparison of results obtained in this work to other determinations~\cite{Braun:2015axa,Bali:2018spj,RQCD:2019osh,Zhang:2020gaj,Detmold:2021qln,Gao:2022vyh,Cloet:2024vbv} of the low Mellin moments of the pion LCDA. 
}
\label{fig:comparison}
\end{figure*}

It is interesting to compare the results of this work with other determinations of the low Mellin moments of the pion LCDA using LQCD. More results exist for the second Mellin moment than the fourth Mellin moment. Other results for both are shown in Fig.~\ref{fig:comparison}. The result presented for the second Mellin moment is  consistent with the previous determination using a subset of the ensembles used here although it is smaller (at several sigma) than other recent results for this quantity. There are several reasons why this may be so. In particular, this calculation was performed at a light quark mass corresponding to a pion mass of $m_\pi\approx 550~\si{MeV}$, while other recent determinations either employ an explicit extrapolation to the physical pion mass~\cite{Braun:2015axa,RQCD:2019osh}, or work directly with light quark masses which produce a physical pion mass~\cite{Gao:2022vyh,Baker:2024zcd,Cloet:2024vbv}. Works that study the quark mass dependence of this quantity, for example, Ref.~\cite{Braun:2015axa}, observe approximately a one percent increase in the second Mellin moment of the pion LCDA as they vary the quark mass from approximately $550~\si{MeV}$ to the physical pion mass. Thus while the slope has the correct sign to reduce the discrepancy between these two results, the magnitude is too small to explain the difference between the result presented here and other determinations of this quantity. 
Of other results shown in Fig.~\ref{fig:comparison}, those from Refs.~\cite{Braun:2015axa,RQCD:2019osh,Zhang:2020gaj,Holligan:2023rex,Detmold:2021qln}
are in the continuum limit. Thus other results contain a systematic error from working at finite lattice spacing which is not estimated.
Finally, in addition to the differences in pion mass and lattice artifacts, the calculation presented here was also performed using the quenched approximation. Empirically, it has been observed that quenching can lead to an uncontrolled systematic error of the order of 10-20 percent, but since there is no way of rigorously estimating this uncertainty without performing the dynamical calculation, here the result is quoted without this source of systematic error. 

The result presented in this article constitutes the first continuum limit determination of the fourth Mellin moment of the pion LCDA. Three previous determinations of the fourth Mellin moment exist in the literature, each obtained at finite lattice spacing. The result presented in this work for the fourth Mellin moment is in agreement with Ref.~\cite{Bali:2018spj}, but smaller than the previous determination in Ref.~\cite{Gao:2022vyh} at a 1 sigma level. However, as with the second Mellin moment, underestimated systematic errors arising from the quenched approximation and the use of an unphysical pion mass (in this work) and lattice artifacts (in other works) prevent a direct comparison of these results.
There exists a larger discrepancy between Ref.~\cite{Cloet:2024vbv} and this work, which was determined using the LaMET framework. Formally, LaMET gives direct access to a range of $\xi\in[\xi_\text{low},\xi_\text{high}]$ where $-1<\xi_\text{low}<\xi_\text{high}<1$. Thus in order to convert the $\xi$-dependence into constraints on the Mellin moments, the endpoints must be modeled. Therefore, the result presented in Ref.~\cite{Cloet:2024vbv} contains an additional systematic uncertainty from the modeling of these endpoints which is not estimated.

At asymtotically large renormalization scales, the form of the LCDA is $\phi(\xi,\mu\to\infty)=\frac{3}{2}(1-\xi^2)$.
The corresponding set of Mellin moments are $\braket{\xi^0}=1$, $\braket{\xi^2}=0.2$ and $\braket{\xi^4}=0.086$. Therefore, while the numerical determination for the second Mellin moment presented here is in agreement with the asymptotic prediction, the fourth Mellin moment disagrees with it at just over one sigma, providing weak evidence that, at the scales considered here, the LCDA has not yet evolved to its asymptotic form. While it is interesting to speculate about the implications of this work for phenomenology, in particular for predictions of the pion electromagnetic form factor, the presence of uncontrolled systematic uncertainties from quenching precludes a detailed analysis. The systematic error incurred from quenching and the use of an unphysical pion mass will be addressed in future work.

\section{Conclusions}\label{sec:conclusions}
The pion light-cone distribution amplitude constitutes a central observable of interest for exclusive processes in QCD. One approach to extracting information about this amplitude is via a calculation of the associated Mellin moments. In this work, the HOPE method was employed to compute the second and fourth Mellin moments of the pion LCDA. The result for the second moment was found to be $\braket{\xi^2}(\mu = 2~\si{GeV}) = 0.202(8)$, which is consistent with the previous determination using the same technique, and after accounting for a 10-20 percent systematic error from quenching, is also in agreement with other calculations in the literature. The fourth Mellin moment of the pion at $m_\pi\approx 550~\si{MeV}$ in the quenched approximation was found to be
\begin{equation}
\braket{\xi^4}(\mu = 2~\si{GeV}) = 0.039(28)(11).
\end{equation}
This value is in tension with the previous two determinations of this quantity,  although systematic effects due to the the differences in lattice artifacts present in other calculations, and the use of the quenched approximation used in this calculation likely explain these differences. This result demonstrates that the HOPE method provides a viable  approach to studying the higher moments of the LCDA and other hadron structure observables which may be related to an OPE. Importantly, this study shows that precision determinations of these low moments are possible using the technique described here with current-era computing resources. Further studies using the HOPE method with gauge field ensembles incorporating dynamical fermions at different quark masses are underway and will allow the systematic errors incurred from the quenched approximation and use of an unphysical pion mass to be removed.  

\section*{Acknowledgements}
WD and RJP are supported in part by the U.S. Department of Energy, Office of Science under grant Contract Number DE-SC0011090 and by the SciDAC5 award DE-SC0023116.  RJP has been supported by the projects CEX2019-000918-M (Unidad de Excelencia “Maria de Maeztu”), PID2020-118758GB-I00, financed by MICIU/AEI/10.13039/501100011033/ and FEDER, UE, as well as by the EU STRONG-2020 project, under the program H2020-INFRAIA-2018-1 Grant Agreement No. 824093. RJP has been supported in part by Simons Foundation grant 994314 (Simons Collaboration on Confinement and QCD Strings). CJDL is supported by 112-2112-M-A49-021-MY3 and 113-2123-M-A49-001-SVP. The authors thank ASRock Rack Inc. for their support of the construction of an Intel Knights Landing cluster at National Yang Ming Chiao Tung University, where the numerical calculations were performed. Help from Balint Joo in tuning Chroma is acknowledged.
The authors thankfully acknowledge the computer resources at MareNostrum and the technical support provided by BSC (RES-FI-2023-1-0030). The authors gratefully acknowledge the support of ASGC (Academia Sinica Grid Computing Center, AS-CFII-114-A11, NSTC(NSTC 113-2740-M-001-007) for provision of computational resources. This document was prepared using the resources of the Fermi National Accelerator Laboratory (Fermilab), a U.S. Department of Energy, Office of Science, Office of High Energy Physics HEP User Facility. Fermilab is managed by Fermi Forward Discovery Group, LLC, acting under Contract No. 89243024CSC000002. This material is based upon work supported by the U.S. Department of Energy, Office of Science, Office of Nuclear Physics through Contract No. DE-AC02-06CH11357.
The research reported in this work made use of computing facilities of the USQCD Collaboration, which are funded by the Office of Science of the U.S. Department of Energy.

\appendix

\section{Fits to HOPE Formula}\label{app:hope_fits}

The hadronic matrix element defined in Eq.~\eqref{eq:data_ratio} is fitted to the one-loop form of the HOPE equation given in Eq.~\eqref{eq:Gegen_OPE_had_amp}. Since the choice of fit window is determined by goodness of fit, fit parameters are resampled as described in Sec.~\ref{sec:numerical_details} to account for the systematic error stemming from the choice of $t_\text{start}$ and $\Delta$. Highest weight fits for each ensemble are shown here.

\begin{figure*}
\centering
\includegraphics[width=0.45\linewidth]{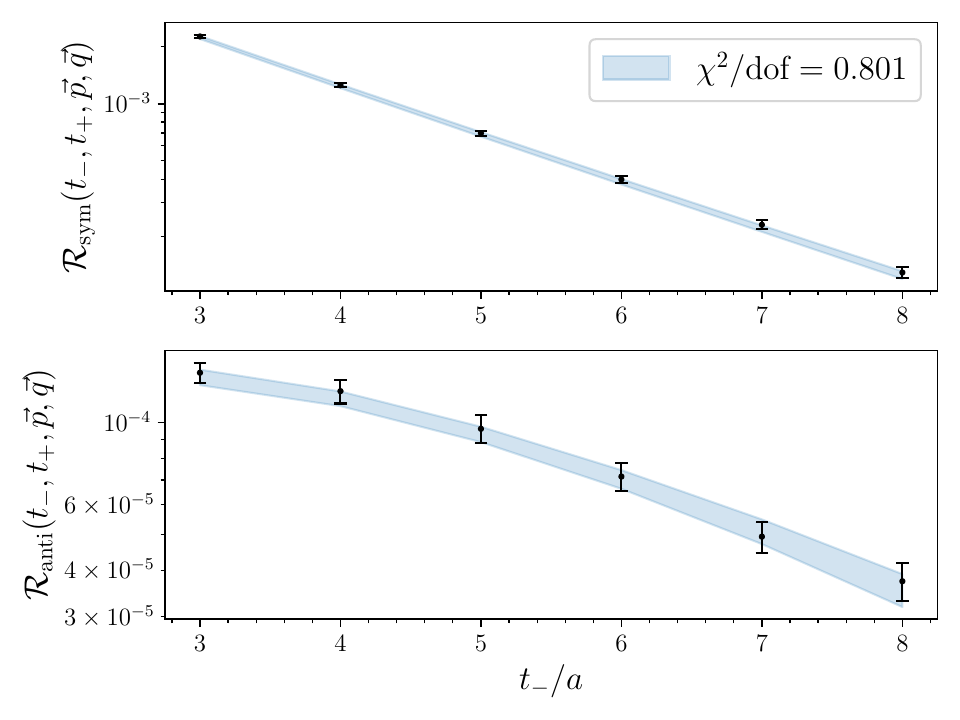}
\includegraphics[width=0.45\linewidth]{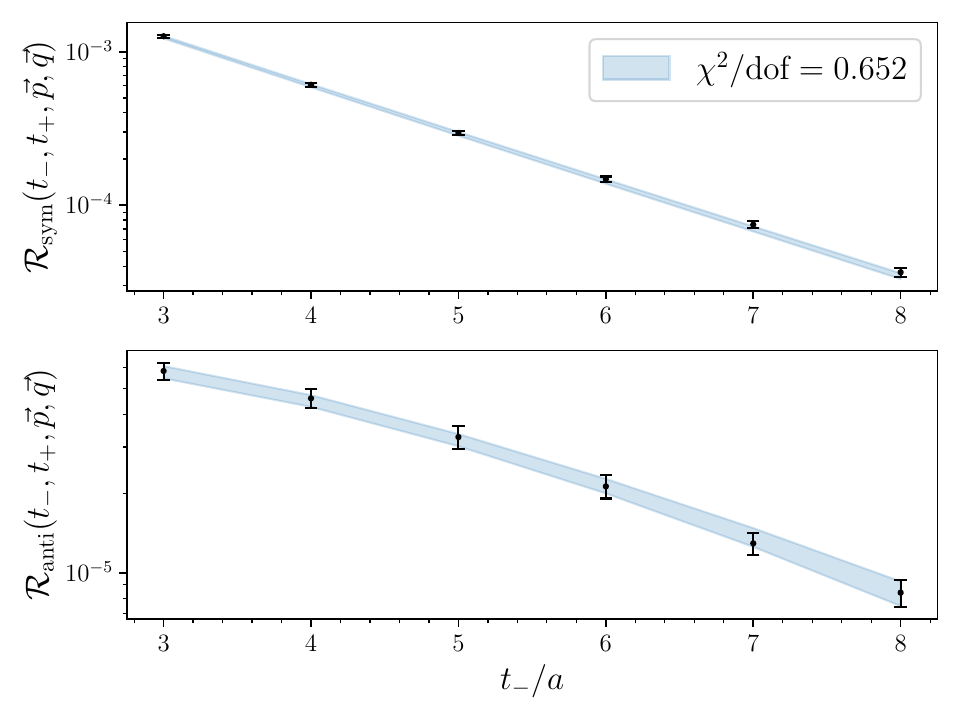}

\includegraphics[width=0.45\linewidth]{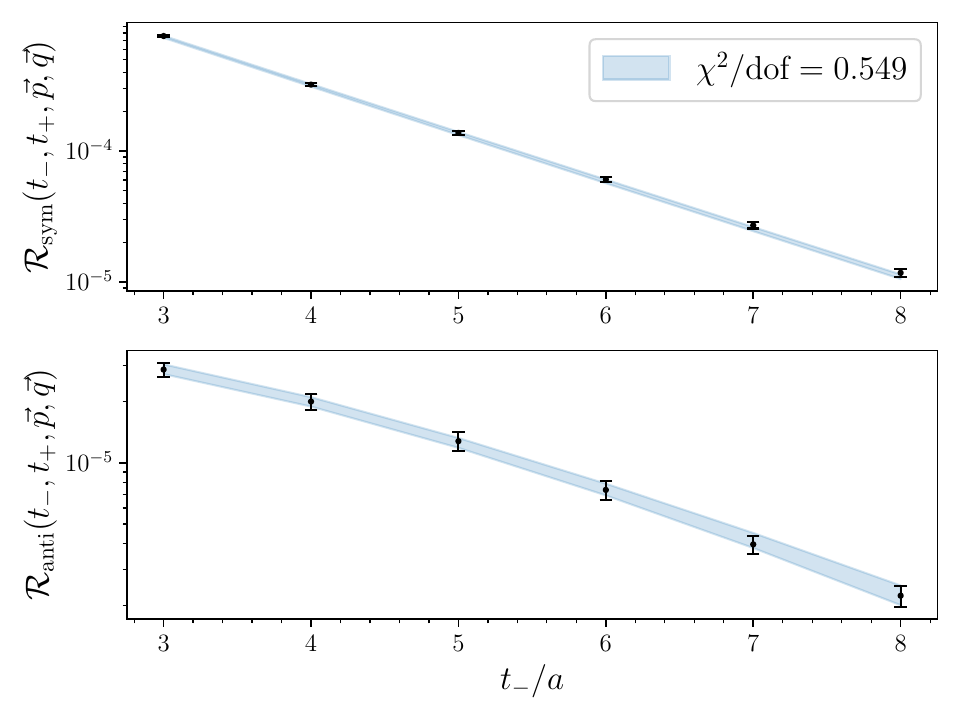}
\includegraphics[width=0.45\linewidth]{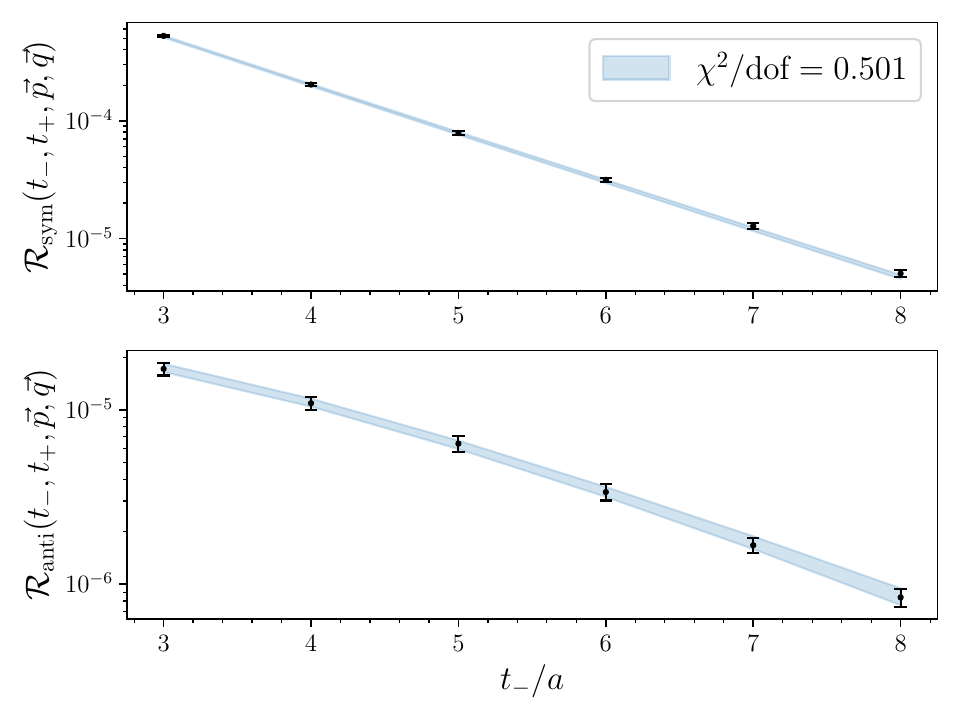}

\includegraphics[width=0.45\linewidth]{ratio_fit_L24_kappa_0.11_tstart_3_tstop_9_mu_2.0.pdf}
\caption{The symmetric and anti-symmetric parts of hadronic matrix element and the resulting best-fit curve with statistical uncertainties obtained by using the HOPE formula given in Eq.~\eqref{eq:Gegen_OPE_had_amp}. The plots correspond to $L/a = 24$, $\kappa_\text{H} = 0.130,~0.125,~0.120,~0.116,~0.110$.}
\label{fig:characteristic_fits_24}
\end{figure*}

\begin{figure*}
\centering
\includegraphics[width=0.45\linewidth]{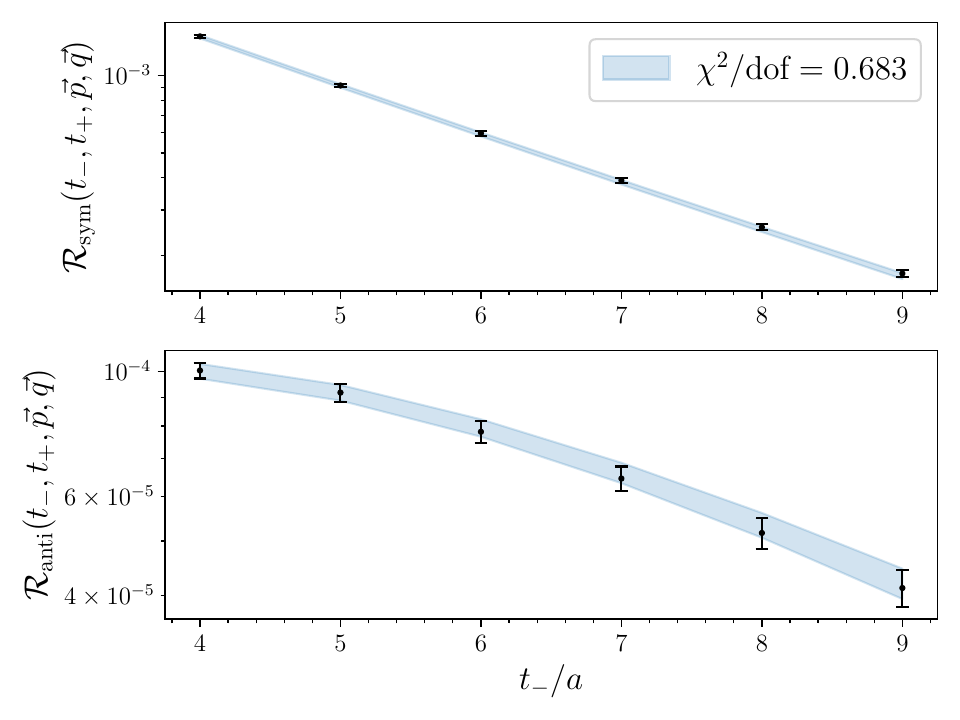}
\includegraphics[width=0.45\linewidth]{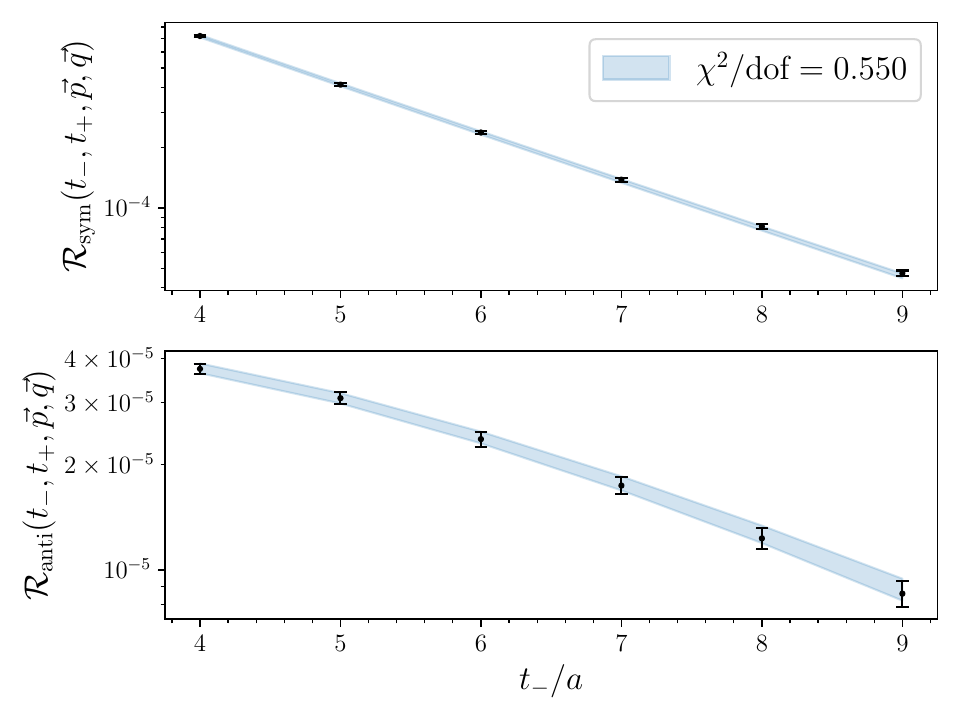}

\includegraphics[width=0.45\linewidth]{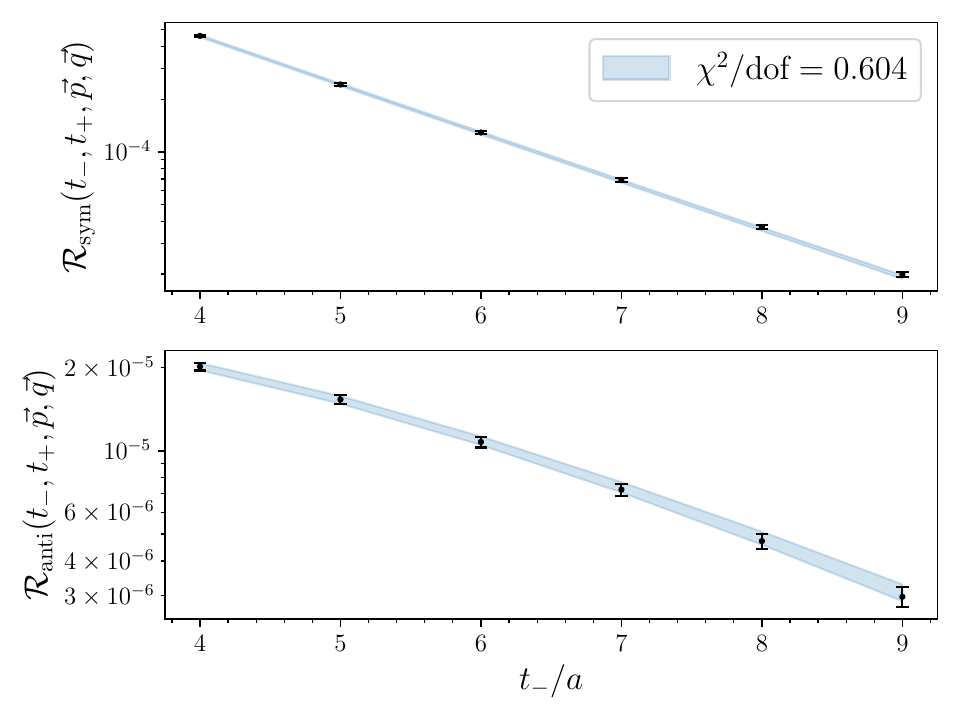}
\includegraphics[width=0.45\linewidth]{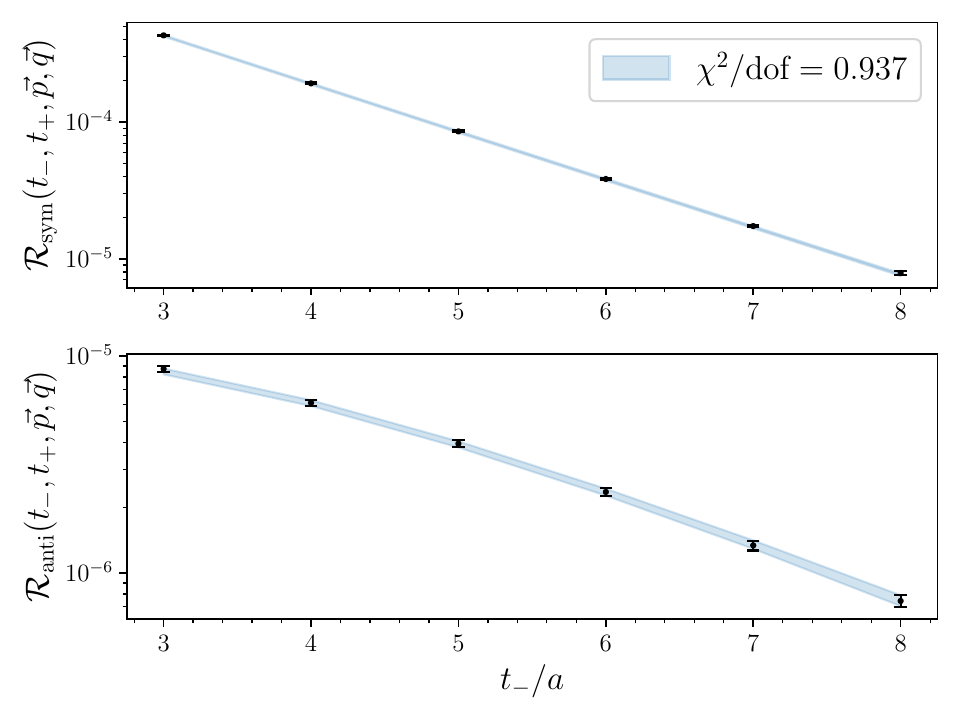}

\includegraphics[width=0.45\linewidth]{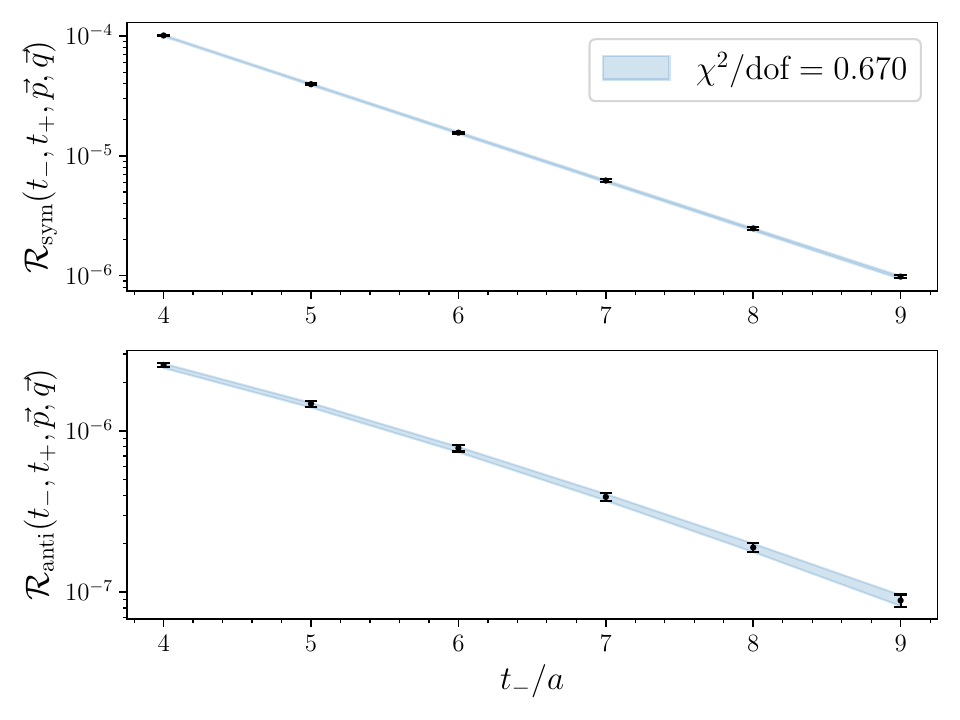}
\includegraphics[width=0.45\linewidth]{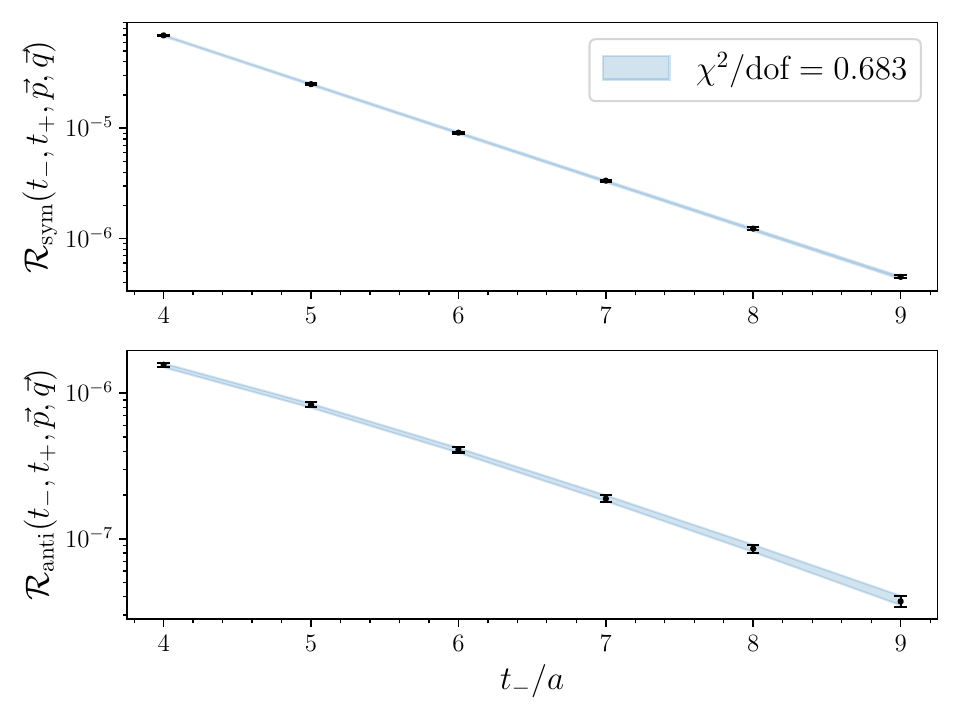}

\caption{Same as in Fig.~\ref{fig:characteristic_fits_24} for $L/a = 32$, $\kappa_\text{H} = 0.1320,~0.1280,~0.1250,~0.1184,~0.1130,~0.1095$.}
\label{fig:characteristic_fits_32}
\end{figure*}

\begin{figure*}
\centering
\includegraphics[width=0.45\linewidth]{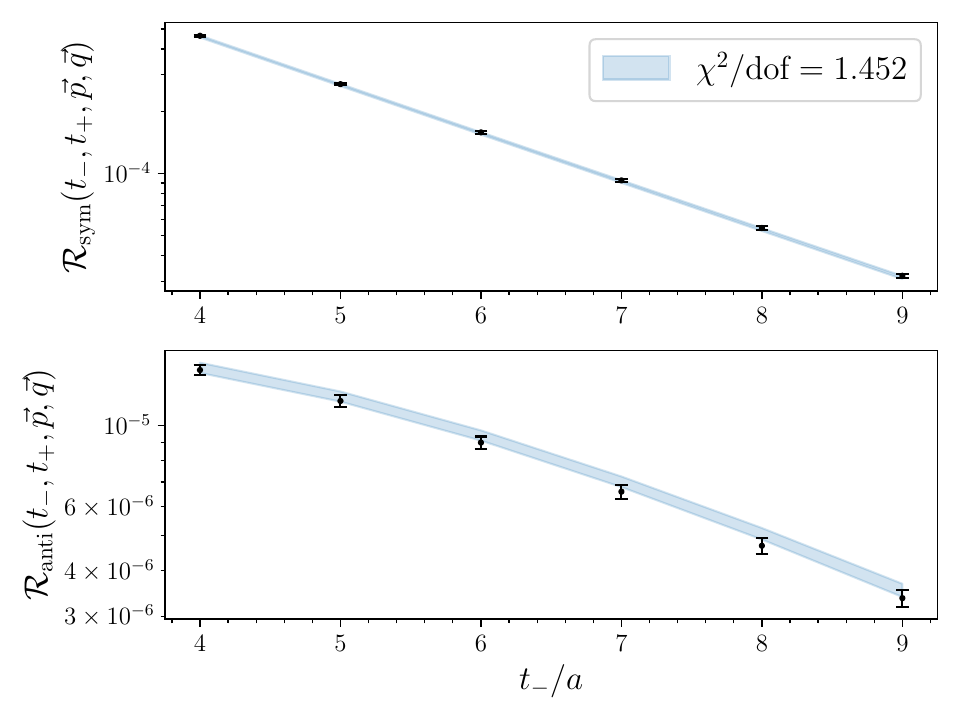}
\includegraphics[width=0.45\linewidth]{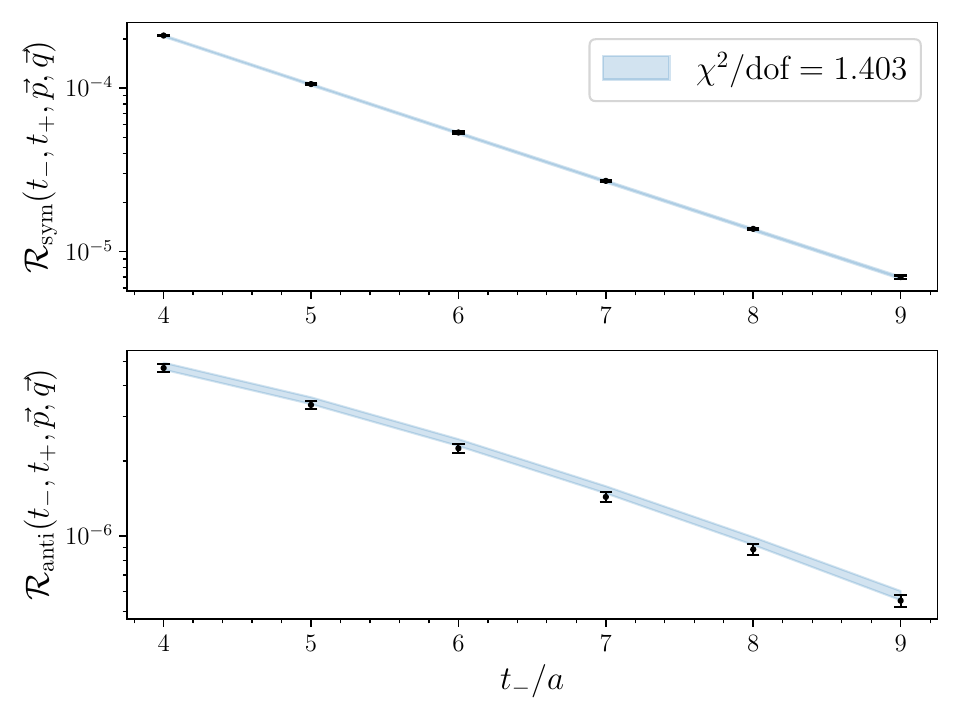}

\includegraphics[width=0.45\linewidth]{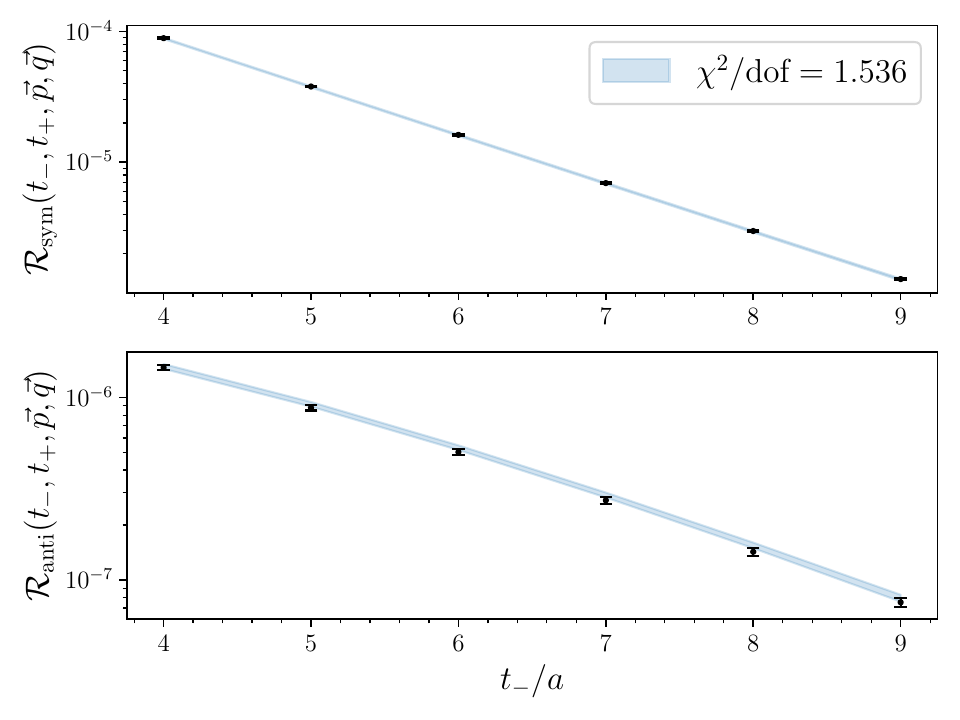}
\caption{Same as in Fig.~\ref{fig:characteristic_fits_24} for $L/a = 40$, $\kappa_\text{H} = 0.1270,~0.1217,~0.1150$.}
\label{fig:characteristic_fits_40}
\end{figure*}

\begin{figure*}
\centering
\includegraphics[width=0.45\linewidth]{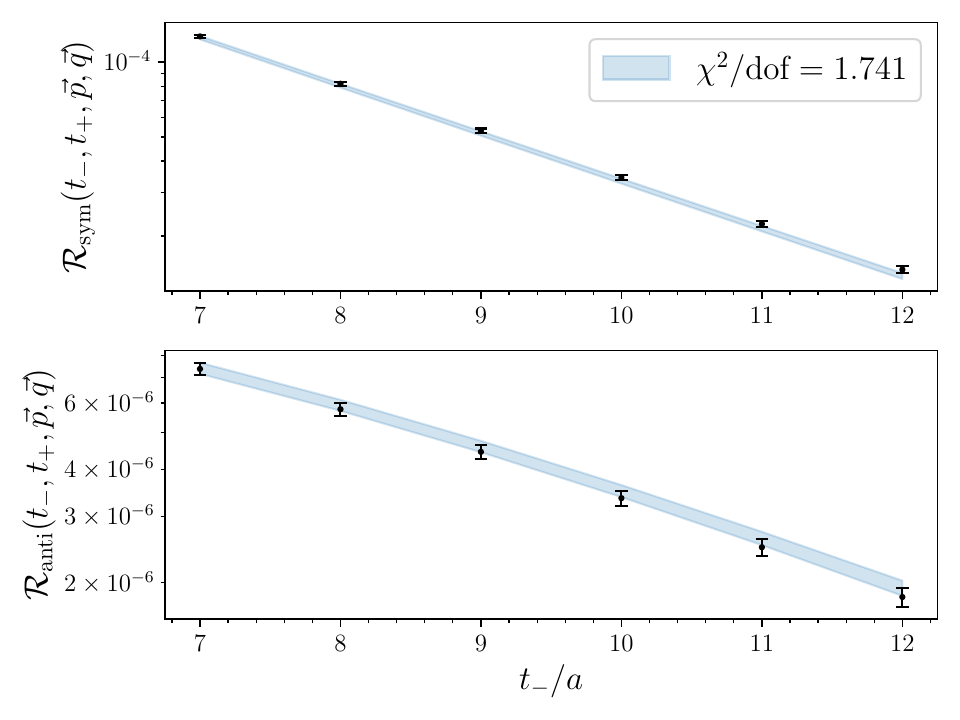}
\includegraphics[width=0.45\linewidth]{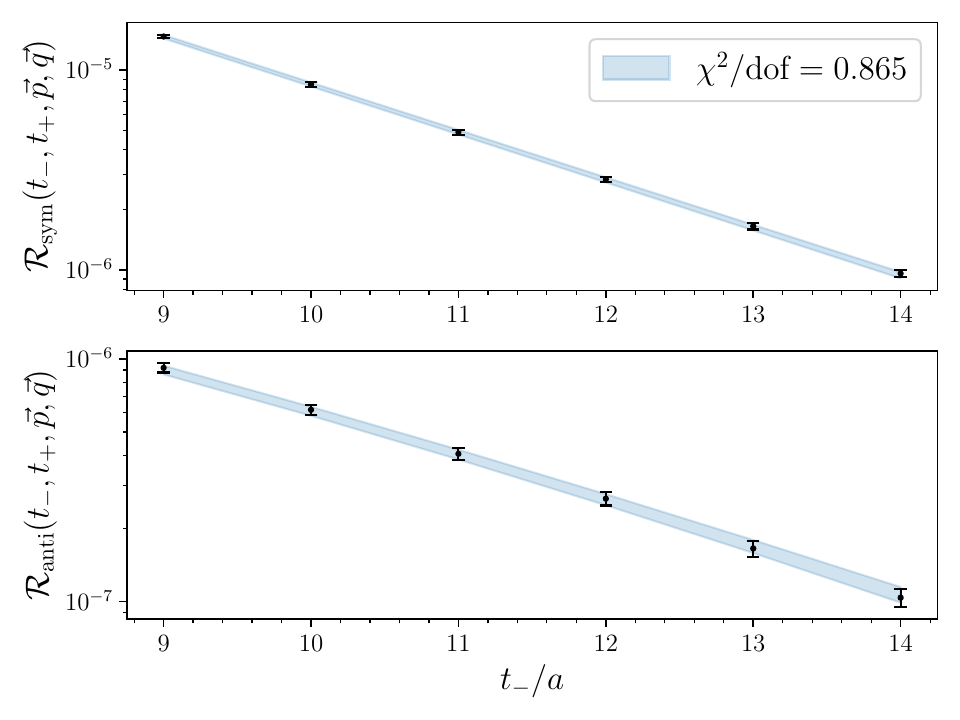}

\includegraphics[width=0.45\linewidth]{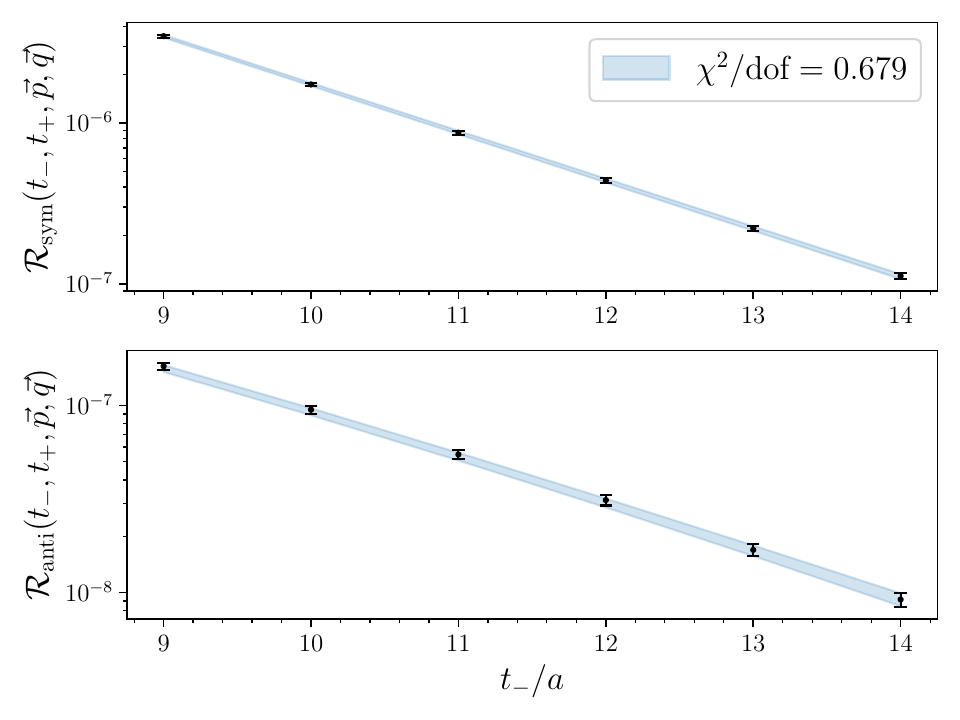}
\includegraphics[width=0.45\linewidth]{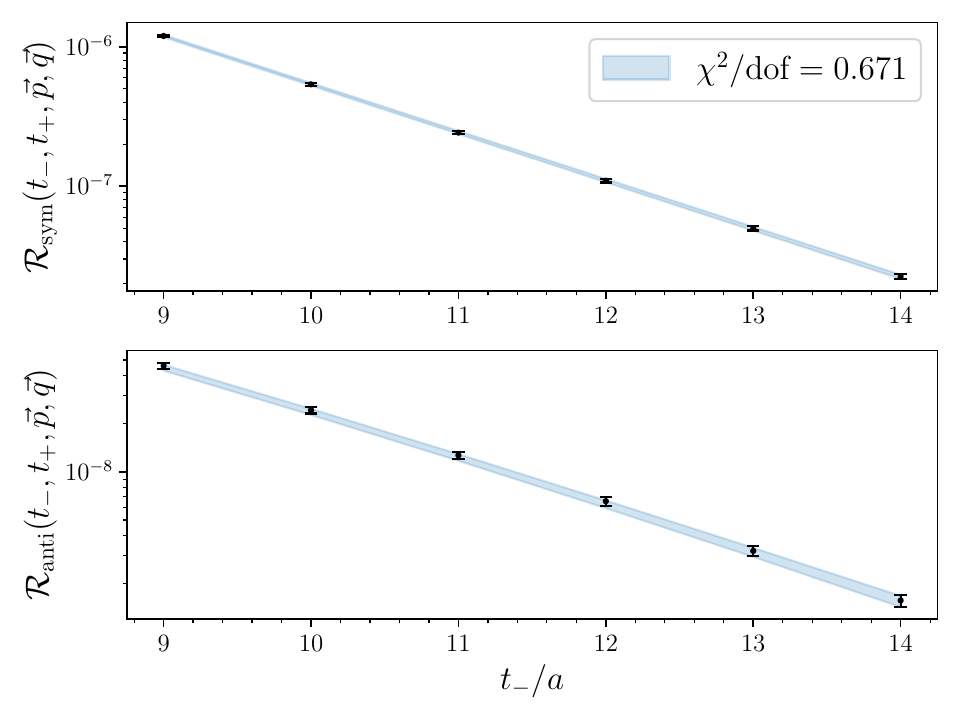}
\caption{Same as in Fig.~\ref{fig:characteristic_fits_24} for $L/a = 48$, $\kappa_\text{H} = 0.1285,~0.1244,~0.1192,~0.1150$.}
\label{fig:characteristic_fits_48}
\end{figure*}

\newpage
\clearpage

\bibliography{refs}

\end{document}